\documentclass{article}
\usepackage{url}
\usepackage{verbatim}
\usepackage{amsmath}
\usepackage{epsfig}
\usepackage{framed}
\usepackage{multicol}

\begin{document}
\title{Topic Diffusion and Emergence of Virality in Social Networks}

\author{
S.Rajyalakshmi
\and
Amitabha Bagchi
\and 
Soham Das 
\and
Rudra M. Tripathy \\
\and
Department of Computer Science \& Engineering \\
Indian Institute of Technology Delhi, New Delhi 
}
\date{}
\maketitle

\begin{abstract}
  We propose a stochastic model for the diffusion of topics entering a
  social network modeled by a Watts-Strogatz graph. Our model sets
  into play an implicit competition between these topics as they vie
  for the attention of users in the network. The dynamics of our model
  are based on notions taken from real-world OSNs like Twitter where
  users either adopt an exogenous topic or copy topics from their
  neighbors leading to endogenous propagation. When instantiated
  correctly, the model achieves a viral regime where a few topics
  garner unusually good response from the network, closely mimicking
  the behavior of real-world OSNs. Our main contribution is our
  description of how clusters of proximate users that have spoken on
  the topic merge to form a large giant component making a topic go
  viral. This demonstrates that it is not weak ties but actually
  strong ties that play a major part in virality. We further validate
  our model and our hypotheses about its behavior by comparing our
  simulation results with the results of a measurement study conducted
  on real data taken from Twitter.
\end{abstract}

{\bf Keywords: }\emph{Social networks, information diffusion, virality}
\section{Introduction}
\label{sec:intro}

Online Social Networks are ubiquitous, as is the realization that our
times are being reshaped by them.  On the more successful of these
platforms, vast user bases distributed around the world are in
continuous conversation, transacting information, thoughts and
ideas--we use the term {\em topic} in this paper--in a variety of
formats. Sometimes these transactions lead an otherwise obscure notion
to global fame. This possibility of a rags-to-riches story in the
world of memes is the wellspring of much of the excitement surrounding
OSNs, an excitement captured in the coinage ``going viral''.
Understanding the processes which make certain topics go viral while
others do not is crucial, and could have wideranging economic, social
and political consequences. In this paper we present and study a model
of topic diffusion on social networks, along with a model-driven study
of a real data set, in order to shed some light on these
processes. Our main insight is that virality on social networks is
{\em not} helped by long range links. Virality, in fact, occurs when
several close knit communities of users get interested in a topic to
such an extent that they merge to form a very large cluster, causing
the popularity of the topic to soar. 

We define a stochastic model whose dynamics are inspired by real user
behavior, and whose macro characteristics closely resemble those of
real OSNs. This model, the first comprehensive attempt of its kind to
our knowledge, could be extremely useful in generating traces of topic
diffusion in OSNs for a multitude of applications like infrastructure
provisioning, viral marketing, disaster management etc, but our goal
in defining this model is more fundamental: We hope that studying the
mechanisms that engender virality in our model will lead to a basic
understanding of the phenomenon of virality in social networks.  

Our model is a time-evolving stochastic model that captures some of
the key features of the diffusion of topics in a social network. A
stream of exogenous topics, we call this the {\em global list}, is
presented to the network's users as global list. The times of
appearance of these topics is given by a Poisson point process. Each
exogenous topic has a weight that decays exponentially with the
passage of time, reflecting the way in which interest in topics
decreases as they get older. The users also perform actions at times
whose distribution follows a Poisson process independent of the time
distribution of other users and of the global list. At each Poisson
time the user decides to talk on a topic picked randomly from the
global list or from one of the topics in the lists of its
neighbors. This action creates an instance of that topic in the user's
list. This instance has a weight that also decays exponentially. The
probability distribution by which the user decides which topic to
speak on is determined by the weights of the topics in the global list
and the weights of instances in its neighbors' lists.

The decay in weights of older topics and older instances of topics
means that the model has an inherent bias towards newer topics. Older
topics retain or grow in strength in local neighborhoods only if those
neighborhoods see a growth in instances being created in the
neighborhood, capturing the notion that users in a social network are
often peer pressured into responding to a particular topic if their
connections are interested in it. Our model is, therefore, implicitly
a competitive model. The competition between topics here is not
explicit, but it is real in a sense that advertisers and political
parties will immediately recognize: It is a competition for the
limited mindspace of the user base. Our model ends up capturing this
notion of a limited mindspace by ensuring that the total weight
of instances resident on a node is stationary (in a probabilistic
sense) i.e. it might be elastic to some extent but on average over a
long period of time it is bounded.

In this study we have used a small-world network to model the social
network graph. The reason for this is that we are focusing on
ordinary users whose connections show small world properties and have
a high degree of reciprocity. The very high impact users that are
present on OSNs like Twitter--news media, celebrities--are not present
in our network. We abstract their influence into the global list. In
doing so we take the focus away from those topics whose wide spread is
due to overwhelming promotion by high impact users and turn the lens
on those topics whose virality is a consequence of the interest shown
in them by the larger network of ordinary users. The small world
network respects constraints on the number of connections a normal
person can maintain similar to those suggested by
Dunbar~\cite{dunbar-jhumanevol:1992}. 

The fundamental contribution of this paper is the model mentioned
above. We describe this in detail in Section~\ref{sec:model}. This
model does not show behavior similar to a real-world OSN for all
settings of its parameters but we do find a parameter regime, we call
it the {\em viral regime}, for which it does. To study this regime we
proceed with a very simple notion of a viral topic: A topic which sees
a very large number of instances created at any point of time is
deemed viral. We use the vague notion ``very large'' to allow us
greater flexibility while conforming to a popular notion of what
constitutes a viral topic.  In Section~\ref{sec:virality} we describe
the characteristics of a viral topic in the viral regime, particularly
emphasizing the emergence of a giant component in the subgraph induced
by users that have spoken on the topic. In Section~\ref{sec:merging}
we discuss how the growth of the topic in several strongly-linked
neighborhoods--we call them {\em lattice clusters}--leads to viral
behavior when these strongly-linked neighborhoods merge, through
strong links (as opposed to weak long-range links), to form a very
large lattice cluster that forms the backbone for the wide spread of
the topic. This narrative of ascent to virality is the main
contribution of our paper.

In Section~\ref{sec:preliminaries} we describe our simulation set-up
and the measurement study whose results we use to validate our
model. We discuss some important strands of the literature in
Section~\ref{sec:intro:related}. Finally, we conclude in
Section~\ref{sec:conclusions} with some discussion of the implications
of our work and future directions.

\subsection{Related Work}
\label{sec:intro:related}
There have been several studies on information diffusion through
social networks, in an attempt to understand the dynamics behind such
networks and explore virality. Leskovec et al.~\cite{meme_tracking}
show how the growth of one topic affects the growth of other topics in
the blogosphere. Yang et al.~\cite{Yang_2011} study the temporal
evolution of topics and show how the popularity of a topic grows and
fades over time.  Sousa et al.~\cite{Sousa_2010} have studied how
topics spread in Twitter. Using three topics: sports, religions and
politics, they investigate the behavior of users in spreading a
topic. In another important work, Khaw et al.~\cite{what_is_twitter}
study topological characteristics of Twitter and analyze how the
trending topics are used by Twitter users. Tao et al.~\cite{analyzing}
conduct a measurement study on Twitter. They observe the evolution of
topic strands, the evolution of new users speaking on a topic and
activity patterns of the users.  Earlier, Guha et al.~\cite{info_diff_thru_blog} 
study the diffusion of information through
weblogs, raising the question of how resonance is achieved, resonance
being used for an unusually good response of the network to an
exogenous topic. Our model aims to address precisely this question of
how the dynamics of the network drive an exogenous topic viral.
Hussain et al.~\cite{blogs_spinning} study the evolution of a viral
topic in the blogosphere and the contribution of different classes of
blogs - elite, top general, top political and tail - towards making
the topic go viral.  Wang et al.~\cite{trends_in_social_media} study a
stochastic model to explain viral trends in Twitter, concluding that
retweeting drives virality. This finding is consistent with our claims
since the act of retweeting is an expression of homophily and our
surmise is that homophilic clusters drive virality.

The relevance of a local density of links in spreading a contagion has
been studied in the past under the header of complex propagation
\cite{centola, complex2}. These models show that when simultaneous
existence of multiple activators is required for a node to get
infected, long range links and randomness can slow down the
propagation. The relevance of a clustered local community for
sustenance has also been propounded by Young~\cite{young_pnas} in the
context of the spread of innovation. Goldenberg
et. al.~\cite{goldenberg-marklett:2001} come to a somewhat
contradictory conclusion, claiming that strong and weak ties are
equally important in product adoption through word-of-mouth.

An important difference between these contagion-flavoured works and
ours is that we operate in a competitive setting where the spread of a
topic limits the spread of others. A similar setting is found in the
competitive viral marketing~\cite{Bharathi_2007, Carnes_2007,
  Tomochi_2005}, where there is competition between different
products. Competitive diffusion is also studied with respect to rumor
and anti-rumor~\cite{Tripathy_2010} and good and bad campaigns~\cite{Budak_2011}.

\section{Preliminaries}
\label{sec:preliminaries}

\subsection{The simulation set up}

Our social network is modeled by a Watts-Strogatz graph~\cite{ws}.  We
begin with a regular lattice ring consisting of $n$ nodes. Every node
has a total degree $k$ and is connected to the $k/2$ nearest nodes on
either side of it. We call these local connections, {\em lattice
  edges}. We then consider each lattice edge $(u,v)$ and rewire the
endpoint $v$ to a random node $w$ with a specific probability, known
as the rewiring probability. The new edge now is $(u,w)$. The graph is
constructed in this manner and is known to have structural properties
lying between that of a regular graph and a random graph.

To ensure that the time-evolution of the system is a pure jump Markov
processes (with the intention of eventually subjecting it to
mean-field analysis~\cite{kurtz}) we assume that the inter-activity
time of each user follows a Poisson Distribution, as do the
inter-arrival times for new topics in the global list. All these
distributions are independent of each other. The inter-activity times
of all users have the same mean, a significant simplification.

We need very high precision floating point numbers to generate
activity times of nodes. For this, we generated random bit strings of
length 100. Each random bit was obtained using the {\tt rand()}
function in C. These bit strings are used to generate high precision
random numbers between $0$ and $1$.

We ran all our simulations on
Intel(R) Xeon(R) CPU X5550 machine with a 4 GB RAM and a 2.67 GHz
Quad-core processor. We simulated at three different network
sizes (1000,10000 and 100000 nodes) for time-window of 1000
units. This simulation completed in approximately 9 seconds on average
for networks of size 1000, in approximately 100 seconds for size 10000
and in about 19 minutes for size 100000.

\subsection{Measurement study}
The data set used in the measurement study in this paper is a part of
the data we have engineered for an ongoing measurement project being
carried out by our group.  We describe the data set and the
methodology only in outline here since the engineering aspects of
preparing such a data set are complex and require a long
treatment. \footnote{In fact a paper describing our methodology in full detail
  has recently been accepted for presentation at a conference. We
  withold a citation because of the double blind constraint but will
  be happy to share these details with the PC chair in confidence if the
  need arises.}

The data set we engineered is a part of the `tweet7' data set which
was crawled by Yang et al.~\cite{Yang_2011}. We use the first three
month's (11 June, 2009 to 31 August, 2009) tweets from this data
set. The `tweet7' data set contains information only about the tweets
by each user, not the social relationships between the
users. Therefore, to build the social graph of these users, we merged
`tweet7' data set with another data set crawled by Kwak et
al.~\cite{what_is_twitter}. They crawled the information on the
followers of almost all Twitter users during the same period as that
of `tweet7' data. The combined data set contains 196,985,580 tweets
and 9,801,062 users.

We used OpenCalais~\cite{opencalais_url}, a text analysis engine, to
identify the topics from the remaining 90\% of tweets. Using hashtags
and OpenCalais, we are able to extract nearly 6.2M topics in 52M
tweets.  From this set of topics, after a laborious multi-stage manual
process whose details we omit here, we picked out a few topics to
compare their evolution with those of the topics in the simulations.
Since our study focuses on topics that go viral due to the larger
network of common users we mainly targeted topics that are being
talked about by such users and not those being initiated and discussed
by users with very high numbers of follows.  For each such topic, we
define a subgraph, consisting of only the nodes that have talked on
this topic anytime during it's evolution. We put a directional edge on
this subgraph from node $u$ to $v$ only if $v$ talked on this topic
after $u$ had talked on it and there was an edge from $u$ to $v$ in
the actual Twitter graph.

\section{A stochastic model}
\label{sec:model}

The model is hinged on the concepts of novelty and
competition. Novelty implies preference for newer topics over older
ones. And competition adds the element of uncertainty to the
dominance by newer topics. Since these two concepts are general in
nature, our model can be applied to different scenarios. We, however,
focus here on the application of the model to diffusion of topics in a
social network.

\paragraph{Network model} The social network is modeled as an
undirected small-world graph each node represents a user and a link
between two nodes represents a possible interaction between the two
users. We consider here an undirected small world network.

The user network for microblogging sites such as Twitter resemble a
scale-free degree distribution with a few users having a large number
of links followed by a fat tail of users with a low degree. Here we
focus primarily on users constituting the fat tail. The users ranking
higher in the degree distribution behave similar to news media,
exposing a large number of users to a topic. These high ranking users
act like sources for the information, their participation in the
social network being distinct from that of the larger masses. The
question of relevance for such ``celebrity'' users is when to break a
news and at what frequency to keep replenishing it with newer tidbits
in order to maximize its spread. The answer to this question is
dependent on understanding how the information spreads through the
masses. We focus here on the masses which resemble a small world
network and are the core group behind making a content go viral.

Microblogging sites such as Twitter have a follower-following
relationship. Celebrities, public figures and news media updaters
enjoy a high follower count. The nature of interaction is that of a
directed link. However when we come to the masses, the nature of the
links with users which are most frequently used are of a friendship
nature~\cite{networks_that_matter}. We focus here on topics such as
movies, politics, sports, amusing videos, gossip etc. which rely
primarily on these friendship links to become the buzz of the town. We
hence consider here an undirected small world graph.

\paragraph{Topic arrival and propagation processes}

When topics come into being, they are included in a {\em global
  list}. These topics could be ones spoken on by the news media,
tweets by celebrities, events occurring in the real world etc. The
inter-arrival times of the topics follow a Poisson point process with
a mean of $\lambda_1$. The presence of a topic in the global list only
announces its availability. When a user picks the topic from the
global list, we term it {\em adoption}. Only when a topic gets
adopted, does it enter the social network. The nodes could also {\em
  copy} a topic from their neighbors. When a neighbor speaks on a
topic, it creates an instance of the topic in its {\em local list}.

The probabilities of adoption and copying depend on weights that we
assign to the topics in the global list and the instances in the local
lists. These weights decay exponentially with time. We define the
weight parameters later in this section, noting here that we consider
a linear model where the weights of $y$ instances put together is just
the sum of the individual weights of the $y$ instances at that time.
In a microblogging network like Twitter, where the users quickly
browse their timeline and retweet or reply to tweets, multiple
instances do not act in order to convince the user to speak on a
topic. The multiplicity rather acts as a repeated reminder, buzzing to
draw attention. The multiple instances thus linearly add up to draw
more attention of the user.

\paragraph{Diffusion Parameters} The model has the following parameters:
\begin{itemize}
\item {\em Topic interarrival time:} $\lambda_1$. New topics enter the
  system by appearing in a {\em global list}. The times of entry are
  given by a Poisson point process with density $\lambda_1$.
\item {\em User activity time:} $\lambda_2$. Each user performs an
  action which involves creating a new instance of a topic that has
  already appeared in the system. The times of activity follow a Poisson
  point process with density $\lambda_2$. The activity times for all
  the users are independent of each other and of the times of topic
  arrival in the global list.
\item {\em Global weight parameters:} $A, \alpha$. We consider here a simple
  model wherein all topics in the global list are adopted with an
  equal weight $A$. As the topic becomes older and people begin losing
  interest in it, the weight begins to decay. The weight decays
  exponentially with time with a decay parameter of $\alpha$.  The
  weight of a topic in the global list decays with time. For a topic
  that appeared at time $t$, the weight at time $t' \geq t$ is
  $Ae^{-\alpha (t-t')}$ where $A$ is a global weight parameter.
\item {\em Local weight parameters:} $B, \beta$. The weight of an
  instance of a topic that has been spoken of by a user decays with
  time. For a topic instance created at time $t$ by a user, the weight
  of the instance at time $t' \geq t$ is $B e^{-\beta (t-t')}$ where
  $b$ is a global weight parameter.
\end{itemize}

When a node is scheduled to perform an activity, it can either adopt a
topic from the global list or copy any of the instances created by any
of its neighbors with appropriate weights. As the instances of a
particular topic in the neighborhood of a node go up, the copy weight
of the topic also goes up. Since a node has only these two options
when scheduled for activity, we normalize these weights to obtain
probabilities. The normalization factor $c(t)$ is the sum of the
adoption and copy weights of all topics for a particular node at a
particular time $t$. 

\paragraph{Probability of instance creation}
In order to simplify the simulation process we observe that our model
is Markov Process with stationary properties in which the sum of
weights in the global list and in each of the local lists converges to
a stationary value. For the normalization constant we denote this
value $c$. Assuming we have achieved the steady state at time $t$, we
determine the value of $c$ by summing over the adoption and copy
weights of all topics in the global list and all instances in the
local lists of the neighbor of a node scheduled to talk at time
$t$. Using a mean-field approximation we calculate 
\begin{equation}
\label{eqn:csat}
 c = \dfrac{\lambda _1 A}{\alpha} + \dfrac{k \lambda _2 B}{\beta}
\end{equation}
at steady state.

We now use a mean-field approach to estimate the probability of a
user creating an instance of a particular topic at time $t$.  Consider
a topic $i$ that appeared in the global list at time $0$. Let $P(t,i)$
represent the probability that a node talks about topic $i$ at time
$t$.
This probability is the sum of the adoption probability for the topic
and the copy probability of the topic in the neighborhood of the
node. The probability of copying from a neighbor is the just the
probability that the neighbor talked on the topic at some time in the
past, multiplied by the copy probability of that instance at present
time. This when integrated over all times and summed over all
neighbors gives the required probability $P(t,i)$.
\begin{equation}
P(t,i) = \frac{A e^{- \alpha t}}{c} + \frac{k \lambda_2}{c}  \int_0^t B e^{- \beta (t-t^{'})}  P(t^{'},i) dt'
\label{eqn:ptiint}
\end{equation}
where $t$ is the present time and $t'$ is the time an instance was
created by a neighbor. We omit the steps involved in solving this
recursive equation due to paucity of space, presenting only the final
answer: 
\begin{equation}
P(t,i)  =  \frac{A}{c} e^{-D_1 t}  +  \dfrac{D_2}{D_1 - \alpha} \left( e^{- \alpha t}  -  e^{- D_1 t} \right),
\label{eqn:ptisolved}
\end{equation}
where $D_1 = \beta - (k \lambda_1 B)/(c)$ and $D_2 = 
(A(\beta - \alpha))/c$.
At this point it appears that the model has essentially been solved
and there is nothing more to say. But, fortunately or unfortunately,
this is not the case. The picture is complicated by the fact that the 
model displays three different regimes with markedly different
behaviors. 

\paragraph{Three regimes of topic diffusion} Searching through the
parameter space we found three different behavior regimes for our
model. We term these {\em sub-viral}, {\em viral} and {\em
  super-viral}. The rest of the paper will be concerned with the viral
regime because it is the only one whose behavior resembles the real
world. But it is instructive to look at the other two as well.

\begin{figure}[htbp]
\centering
\epsfig{file = 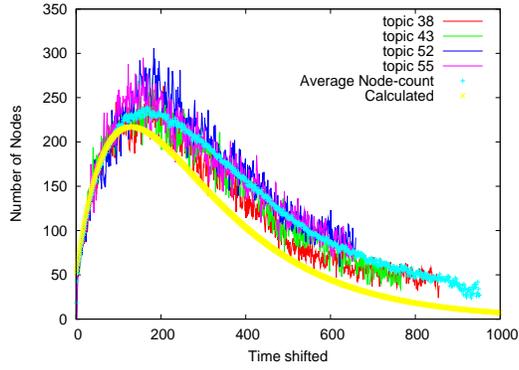, width=7cm,height=5cm}
\caption {Mean-field calculation versus simulated evolutions for the
  sub-viral regime.}
\label{fig:sub_crit_evolution}
\end{figure}

In the sub-viral regime, all topics evolve in a similar fashion. There
is no apparent dominance of any topic over others. The fluctuations in
the peak heights are small enough to fall within the spectrum of
stochastic variances. The network in the sub-viral regime witnesses no
virality. It is in this regime and this regime alone that the
probability value calculate in~(\ref{eqn:ptisolved}) makes sense (see
Figure~\ref{fig:sub_crit_evolution}). The calculated value
from~(\ref{eqn:ptisolved}) and the evolution curves of various topics
(time-shifted to appear to originate at 0 for the plot) show marked
similarity. The efficacy of mean-field analysis in characterizing the
evolution in this regime indicates that long-range interactions play no
part here.
\begin{figure}[htbp]
\centering
\epsfig{file = 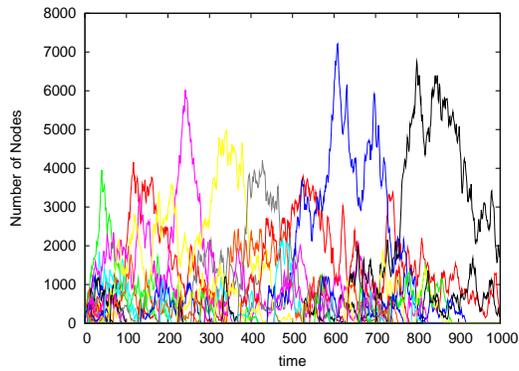, width=7cm,height=5cm}
\caption {Evolution in the viral regime (10000 nodes)}
\label{fig:regimes}
\end{figure}
In the viral regime the mean-field analysis breaks down
completely. This regime is characterized by a marked diversity in the
peak heights for various topics. Figure~\ref{fig:regimes} shows the
evolution of topics in this regime. The fluctuations in the peak
heights are much larger than mere stochastic variances as seen in the
sub-viral regime. The marked distinction in peak heights for some
topics allows us to adjudge these topics viral. We do not put forth
any concrete threshold beyond which a topic may be called viral and
below which non-viral. Any unusually high peaked topic which can be
seen to clearly overpower other topics during its lifetime is taken to
be viral and we study such topics for viral characteristics. Thus, the
viral regime witnesses a self-attained virality as a result of
dynamics through aptly tuned parameters.

Moving further in the same direction in the parameter space, we
observe the end of the viral regime, marked by increasing similarity
in the peak heights for evolution of various topics. Although the
evolution pattern resembles the sub-viral regime, the peak heights are
much higher. The dynamics behind the pattern in this regime is
different from those of the sub-viral regime, making us classify it as
a regime of its own which we call the super-viral regime, and
completely ignore for the rest of this paper since it does not show
real world-like behavior. We now turn our focus to the regime of
interest, the viral regime, beginning by reporting its characteristics
and going on to explain what makes certain topics go viral.

\section{Characteristics of the viral \\ regime}
\label{sec:virality}

\begin{figure}[htbp]
\centering
\epsfig{file = 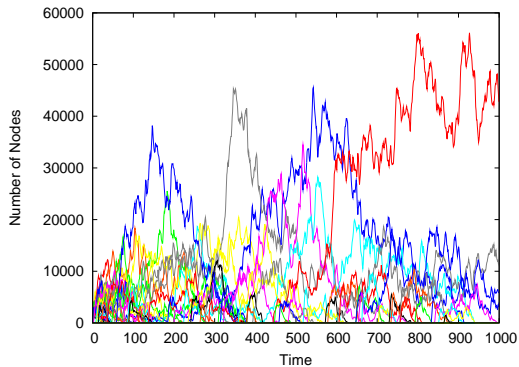, width=7cm,height=5cm}
\caption {Evolution of topics in the viral regime}
\label{fig:viral_evolution}
\end{figure}

The viral regime has a definite diversity in the peaks achieved by
various topics during their evolution. Some topics clearly shoot up
much higher than the average peak of other topics. Such topics which
have an unusually large number of nodes talking about them
simultaneously, are termed viral. Figure \ref{fig:viral_evolution}
shows an example evolution of topics in the viral regime for a
networks of size 100000.
In this section we present some statistical and topological studies
performed on simulation traces of our model when parametrized in the
viral regime. As expected, topic spread, lifetime and peaks all follow
power law distributions. We also study the effect of the number of
topic adopters on the spread of topics and find it to be negligible,
which is what we wanted to achieve.  On the topological front we find
that viral topics have one large connected component while non-viral
topics tend to be highly disconnected. We also present some
topological findings that lay the groundwork for our theories
regarding the emergence of virality (presented in
Section~\ref{sec:merging}.) In this and the subsequent section we say
that a node speaks on topic $i$ at time $t$ is the most recently
created instance in the local list of the node at time $t$ is an
instance of topic $i$.

\subsection{Statistical observations}
\label{sec:virality:observation}
\begin{figure}[htbp]
\centering
\epsfig{file = 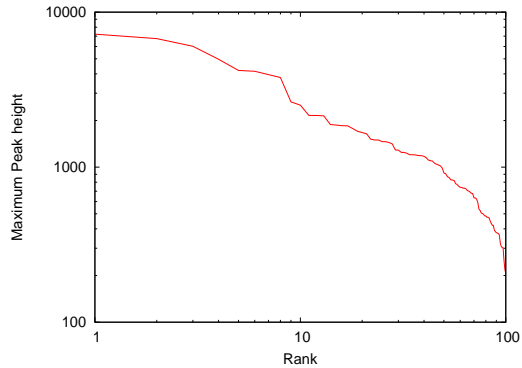, width=7cm,height=5cm}
\caption {Rank ordered plot for maximum peak height of topics. Slope observed is -0.77}
\label{fig:rank_plot}
\end{figure}

\begin{figure}[htbp]
\centering
\epsfig{file = 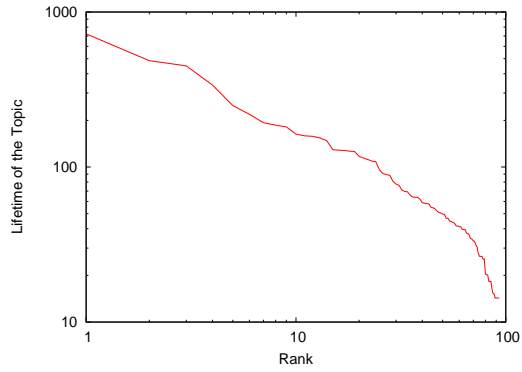, width=7cm,height=5cm}
\caption {Rank ordered plot for lifetimes of different topics. It follows a power law with a slope of -0.7}
\label{fig:lifetime_rank}
\end{figure}

Figure~\ref{fig:rank_plot} shows the rank ordered plot on a log-log
scale for the topics in viral regime, with the topic having the
highest peak during its evolution ranked first. The distribution
follows a power law with very few topics achieving a high peak during
their evolution followed by a fat tail of topics with low peaks.
Figure \ref{fig:lifetime_rank} shows the rank ordered distribution for
the lifetimes of various topics in the network. A topic is said to
survive if there is even a single node in the network speaking on it.
The distribution follows a power law. A few topics live in the network
for a long duration compared to a fat tail of topics which die out
sooner.
\begin{figure}[htbp]
\centering
\epsfig{file = 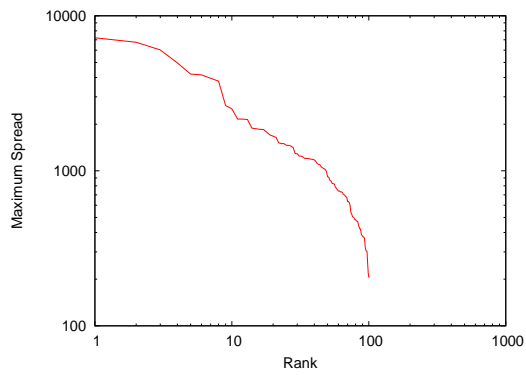, width=7cm,height=5cm}
\caption {Rank ordered plot for maximum spreads for different topics. It follows a power law with a slope of -0.7}
\label{fig:maxspread_rank}
\end{figure}
Figure \ref{fig:maxspread_rank} shows the rank ordered
distribution for the maximum spread of a topic. The maximum spread
depicts the number of nodes which speak about the topic at some point
of time or the other during the evolution of the topic. The power law
indicates that very few topics succeed to reach a high spread while
many topics die out after reaching only a small number of nodes.
\begin{figure}[htbp]
\centering
\epsfig{file = 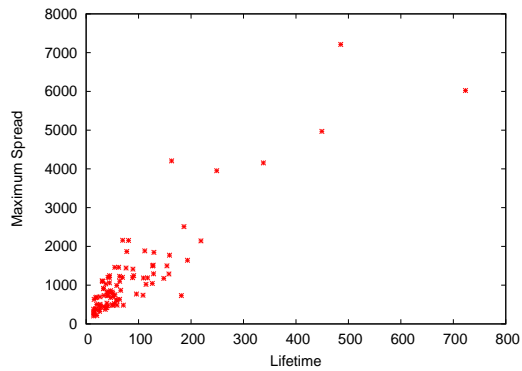, width=7cm,height=5cm}
\caption {Lifetime vs Maximum spread for different topics}
\label{fig:lifetime_maxspread}
\end{figure}
The lifetime vs maximum spread is shown in Figure
\ref{fig:lifetime_maxspread}. It supports the deduction that topics
which live longer in the network are also able to reach out to a
larger section of the network.
\begin{figure}
\centering
\epsfig{file = 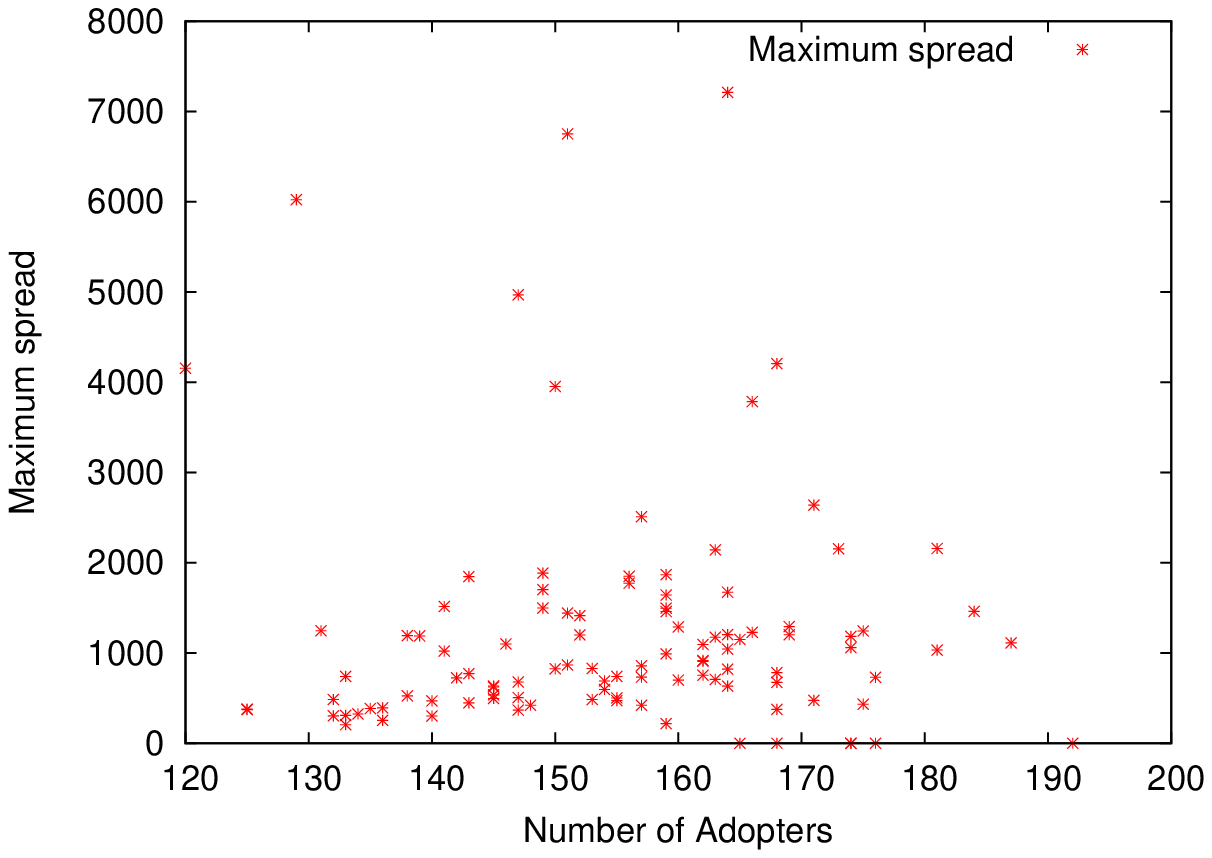, width=7cm,height=5cm}
\caption {Distribution of maximum spread with number of adopters of the topic}
\label{fig:maxspread_adopters}
\end{figure}
\begin{figure}
\centering
\epsfig{file = 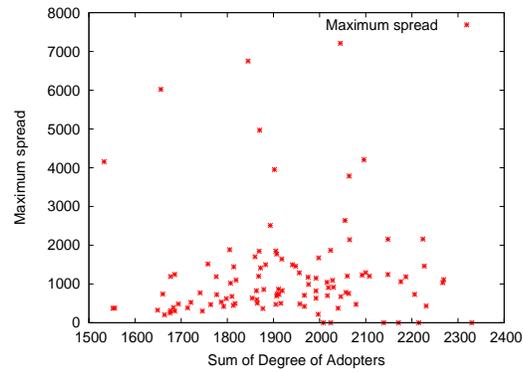, width=7cm,height=5cm}
\caption {Distribution of maximum spread with sum of degrees of the adopters of the topic}
\label{fig:maxspread_sumdegree}
\end{figure}

\subsection{The role of adopters}
\label{sec:virality:adopters}
The role of number of adopters has been crucial in the study of
contagion over the years. The occurrence of high peaks in the
evolution of some topics in the viral regime raises the question of
whether these topics enjoy a higher number of adopters. In our
model, adopters of the topic are the nodes which adopt the topic
from the global list. Since these are the nodes which bring the topic
into the network, one might be tempted to conclude that the number of
adopters for viral topics is higher than that for the non-viral ones.
Figure \ref{fig:maxspread_adopters} shows the distribution of the
maximum spreads achieved by topics along with the number of adopters
of the topic. There is no specific trend in the distribution which can
lead us to believe that the topics enjoying higher number of adopters
also reach out to a larger fraction of the network during their
evolution. Since variances in the degree of the adopters could
significantly affect their reach and hence prove crucial, we
incorporate them before correlating with the spread. Figure
\ref{fig:maxspread_sumdegree} shows the distribution of the maximum
spread of the topics with the sum of the degrees of the nodes which
adopted these topics. We again see no particular correlation and can
thus safely conclude that a higher number of adopters is not the
reason for the peaks observed in the topic evolutions in the viral
regime of our model. This demonstrates that our model does indeed
eschew the effect of high impact users. The topics that go viral in
our setting are the ones that the network itself promotes.

\subsection{Topological characteristics}
\label{sec:virality:topology}
\begin{figure}[htbp]
\centering
\epsfig{file = 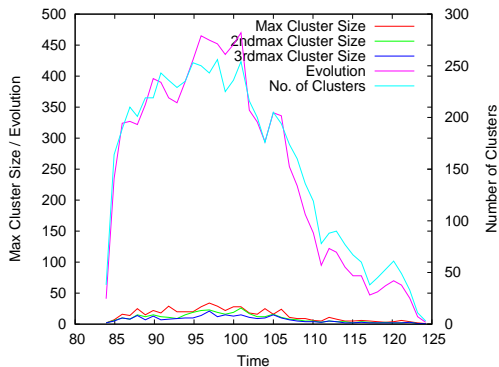,width=7cm,height=5cm}
\caption{Evolution of clusters of nodes speaking on a topic: Non-Viral}
\label{fig:sim_cluster-1}
\end{figure}

\begin{figure}[htbp]
\centering
\epsfig{file = 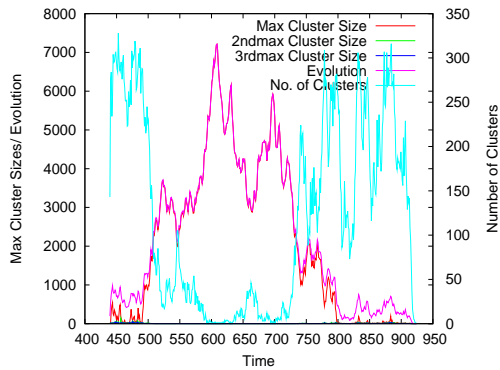, width=7cm,height=5cm}
\caption{Evolution of clusters of nodes speaking on a topic: Viral}
\label{fig:sim_cluster-2}
\end{figure}

\begin{figure}[htbp]
\centering
\epsfig{file = 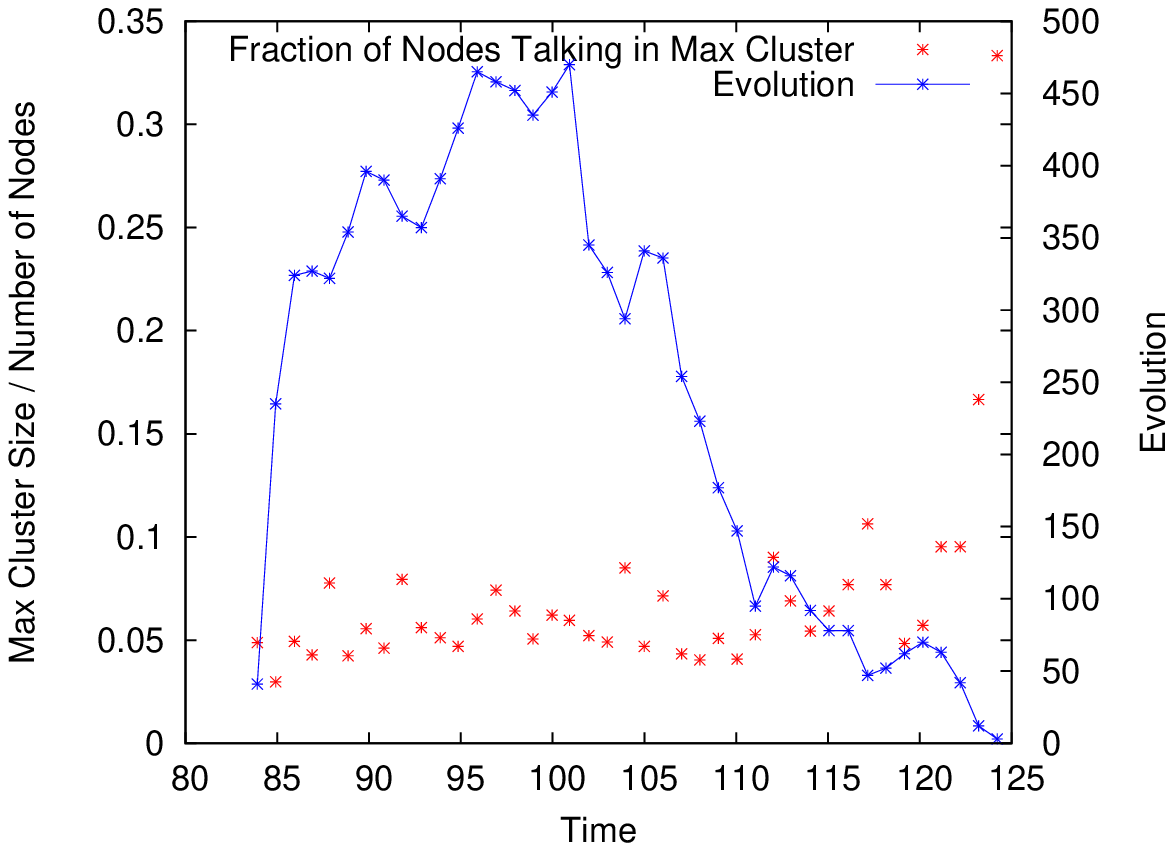,width=7cm,height=5cm}
\caption{Evolution of fraction of nodes lying inside the largest
  cluster: Non-viral}
\label{sim_cluster_fraction-1}
\end{figure}

\begin{figure}[htbp]
\centering
\epsfig{file = 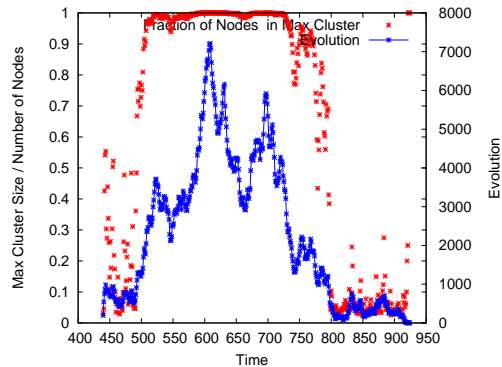, width=7cm,height=5cm}
\caption{Evolution of fraction of nodes lying inside the largest
  cluster: Viral}
\label{sim_cluster_fraction-2}
\end{figure}

To study how the graph topology affects the evolution of the topic, we
look at the evolution of clusters of nodes speaking on a particular
topic simultaneously. Figures~\ref{fig:sim_cluster-1}
and~\ref{fig:sim_cluster-2} show the evolution of the cluster sizes
along with the evolution of the topic for a viral and a non-viral
topic.
A non-viral topic is characterized by many small clusters
disconnected throughout the lifetime of the topic. The number of
clusters increase when there is a peak in the evolution of the topic
indicating that the surplus nodes speaking on the topic belong to
clusters different from the existent ones. On the contrary, for a
viral topic, we see a significant dip in the number of clusters when
the topic peaks in its evolution. This is because of the formation of
a giant cluster which encompasses a large fraction of the nodes
speaking on the topic. The evolution sees similar sized clusters when
the topic is young and the formation of a giant cluster when the topic
peaks. Figures~\ref{sim_cluster_fraction-1}
and~\ref{sim_cluster_fraction-2} show the evolution of the fraction of
nodes lying within the largest cluster. The non-viral sees a very
small fraction inside the largest cluster while the viral sees a large
fraction when it peaks in its evolution indicating that it is a giant
component.


\begin{figure*}[t]
\centering
\begin{tabular}{cc}
\begin{minipage}{5.8cm}
\epsfig{file = 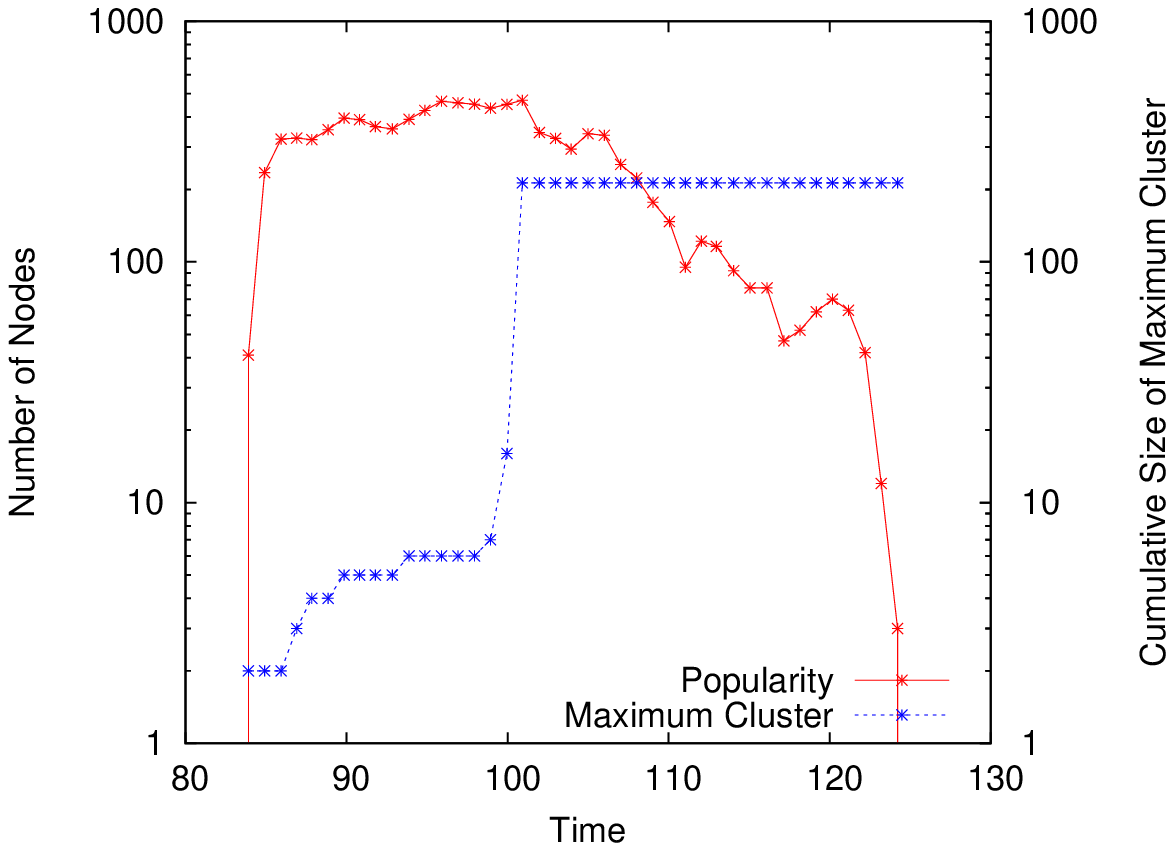,width=5.8cm,height=4cm}
\end{minipage}&
\begin{minipage}{5.8cm}
\epsfig{file = 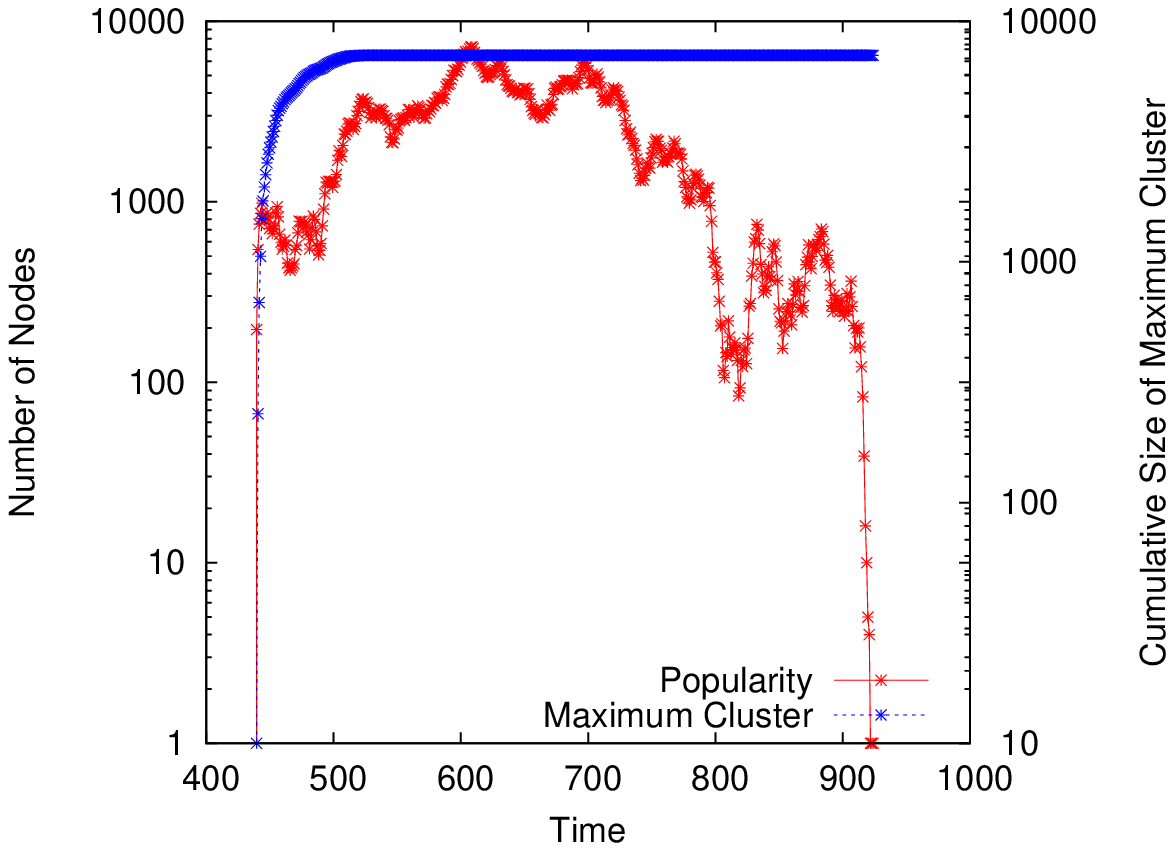, width=5.8cm,height=4cm}
\end{minipage}\\
(a)Simulation: Non-Viral &(b)Simulation: Viral\\
\begin{minipage}{5.8cm}
\epsfig{file = 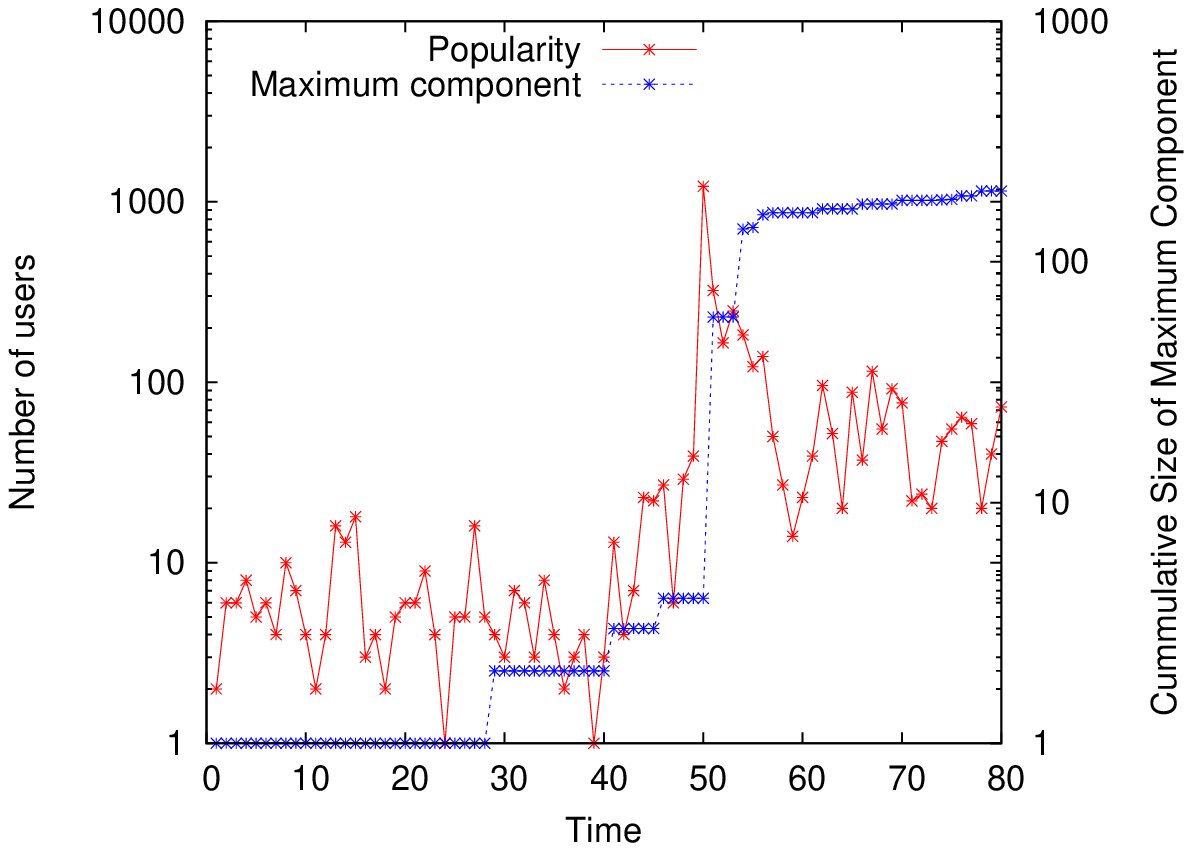, width=5.8cm,height=4cm}
\end{minipage}&
\begin{minipage}{5.8cm}
\epsfig{file = 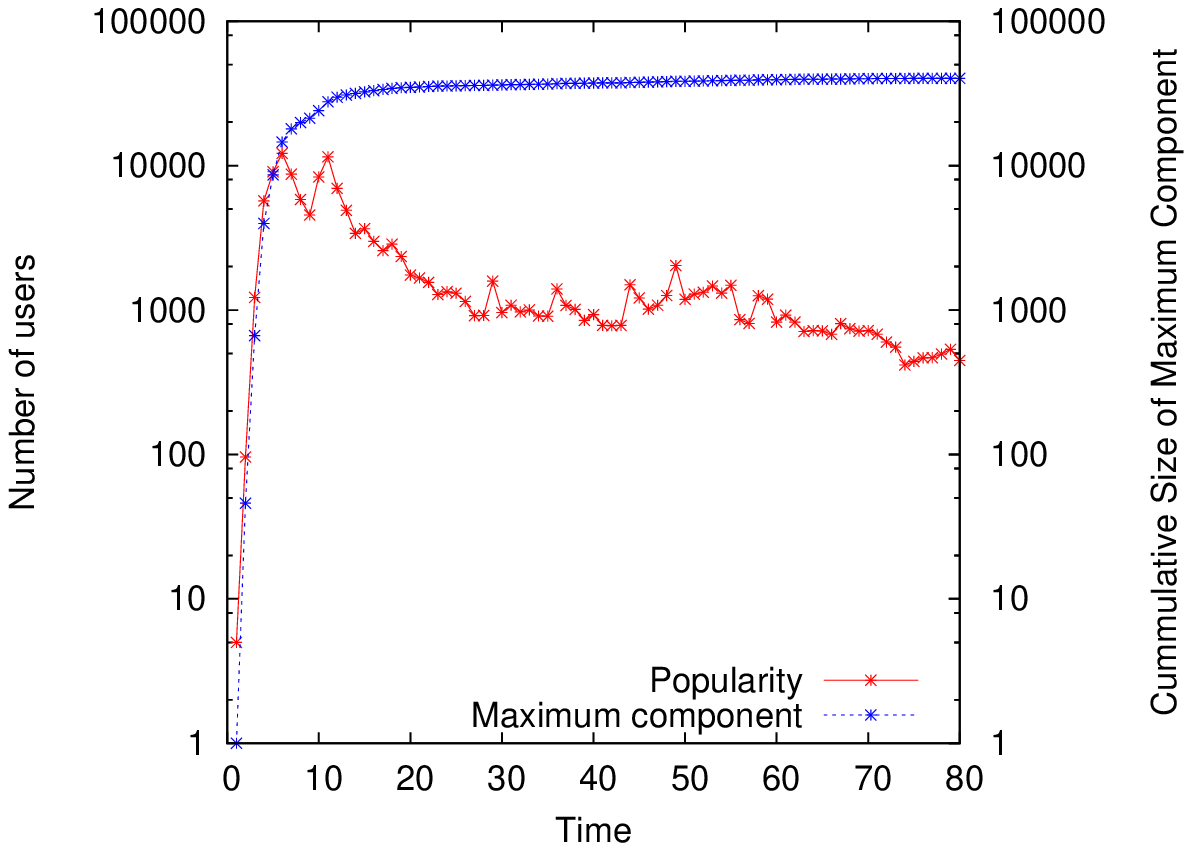, width=5.8cm,height=4cm}
\end{minipage}\\
(c) Real topic: Non-Viral &(d) Real topic: Viral\\
(NICK CANNON) & (IRANELECTION)
\end{tabular}
\caption{Cluster size vs Component size}
\label{fig:component}
\end{figure*}

While the nodes speaking on the viral topic during its peak belong to
a giant component, it is of interest to investigate whether the nodes
have spoken before on the topic or not. In other words, we look at how
the number of unique nodes speaking on the topic in the largest
cluster evolve with time. We term this as the cumulative size of the
largest cluster. In Figure~\ref{fig:component} we see that for a
non-viral topic, the cumulative size grows and saturates at a low
value some time during the evolution of the topic. For a viral topic,
however, we see a sharp increase in the cumulative cluster size as the
topic shoots up. The following times observe a saturation in the
cumulative cluster size while the topic sustains a high peak. This is
representative of the way a viral topic grows. The rise of the topic
is marked by more and more new nodes speaking on the topic while the
sustenance is marked by subsequent discussion among the nodes that
have already talked on the topic. This behavior is also observed in
real topics taken from the Twitter data set.

\begin{figure}[htbp]
\centering
\epsfig{file = 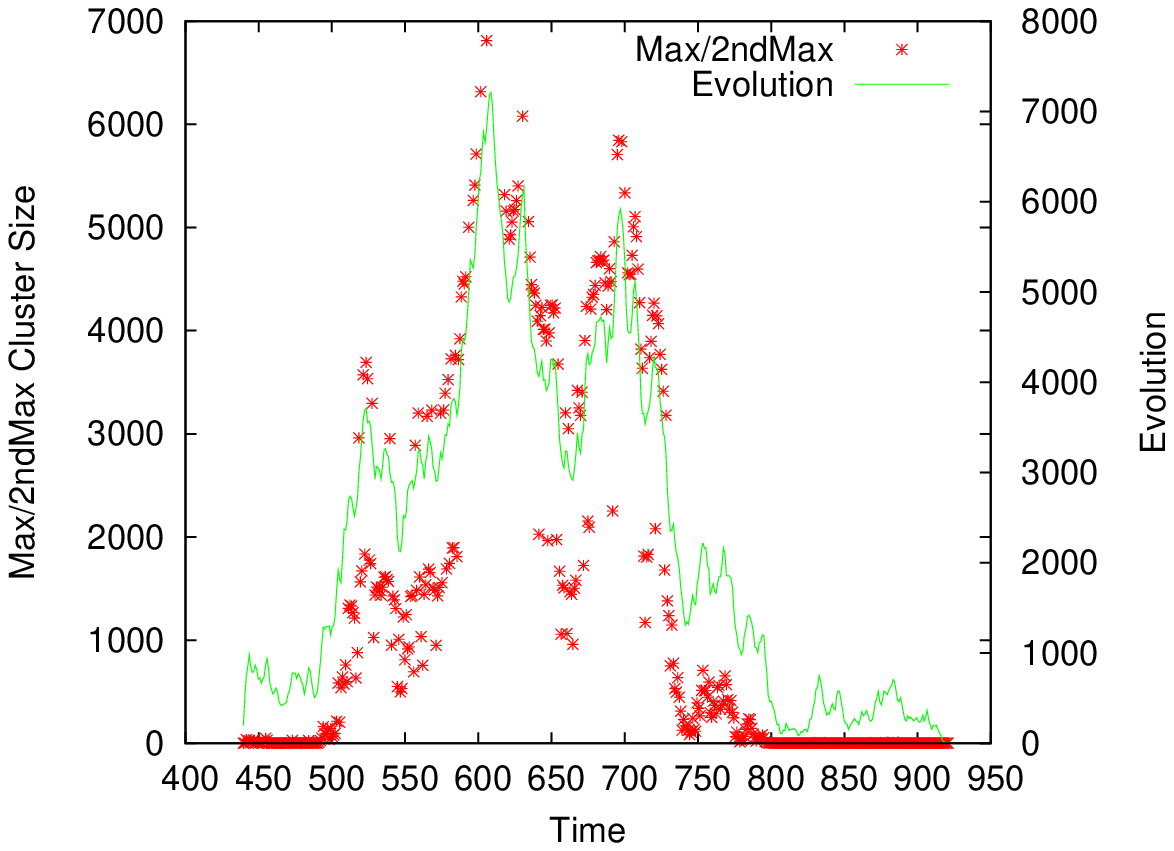, width=7cm,height=5cm}
\caption{Evolution of the ratio of largest to second largest cluster: Viral}
\label{fig:sim_maxby2ndmax-2}
\end{figure}

\begin{figure}
\centering
\epsfig{file = 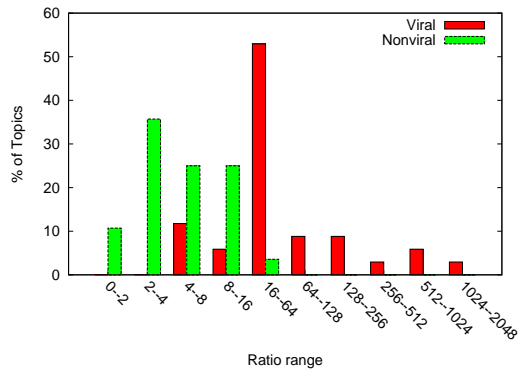, width=7cm,height=5cm}
\caption {Comprehensive histogram depicting ratio of largest to second largest cluster size}
\label{fig:median_ratio}
\end{figure}

For a similar investigation on the real Twitter data set, we look for
the maximum strongly connected component and compare them with the
evolution of the topics. The graphs obtained from the real data bear a
strong resemblance with the ones obtained from the simulations. The
non-viral topic of NICK CANNON sees the cumulative component size
saturating somewhere during its evolution while the one for the viral
topic IRANELECTION saturates as soon as the topic attains its peak.
The topic is then sustained by discussion among these users over the
next several days.
\begin{figure}[htbp]
\centering
\epsfig{file = 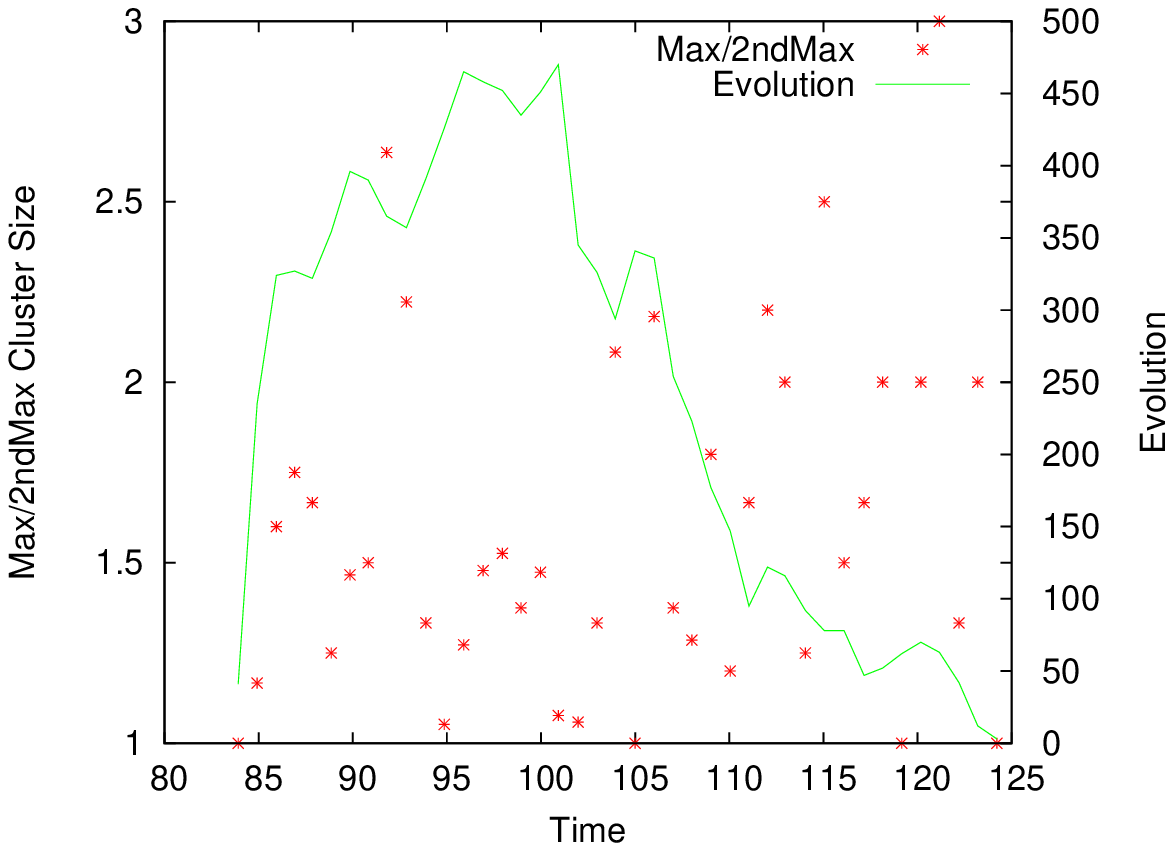,width=7cm,height=5cm}
\caption{Evolution of the ratio of largest to second largest cluster: Non-Viral}
\label{fig:sim_maxby2ndmax-1}
\end{figure}
Having established the significance of a giant component in the
evolution of a viral topic, we represent its formation by plotting the
ratio of the size of the largest cluster to that of the second largest
cluster in Figures~\ref{fig:sim_maxby2ndmax-1}
and~\ref{fig:sim_maxby2ndmax-2}. For a non-viral topic, the ratio does
not fall into any particular pattern and retains low values throughout
the evolution of the topic, thus depicting the lack of formation of a
giant component for a non-viral topic. For the viral topic, the ratios
not only have a higher value but also peak in sync with the peak in
the topic evolution, clearly indicating the formation of the giant
component during the peak and sustenance of a viral topic.
To establish this difference in the giant component formation between
viral and non-viral topics in a comprehensive fashion, we present the
following histogram. We consider the ratio of the largest to the
second largest cluster size and find its median for all topics during
their lifetime. In Figure~\ref{fig:median_ratio}, we plot a histogram for
the different ranges of the median obtained. In the lower ranges of
the ratio, the non-viral topics dominate over the viral ones. Since a
low ratio implies the lack of a giant component, this is an indication
of all non-viral topics lacking such a component. The higher ranges
for the ratio see taller bars for the viral topic, thereby showing the
presence of a giant component in majority of the viral topics.
It may be argued that it is not surprising that if a large number of
users are speaking on a topic the high clustering coefficient of a
Watts-Strogatz graph will make a giant component out of them. However,
the topology of the Watts-Strogatz graph is not the hero (or villain)
of this piece. In fact, we find that even when the long range
(rewired) links of the Watts Strogatz graph are snapped, the largest
cluster of a viral topic constitutes a significant fraction of the
evolution when it takes it high peaks (see
Figure~\ref{fig:lattice_cluster_evolution}). This insight drives the
argument for the emergence of virality that we present in the next
section.

\section{The emergence of virality}
\label{sec:merging}

We now present our narrative of the emergence of virality. In what
follows we use the term {\em local community}. This means those users
who are joined by lattice edges i.e. those edges of the Watts-Strogatz
graph that {\em did not} get rewired to a random node. These local
communities play a critical role in our theory. There are three
crucial hypotheses making up our narrative.
\begin{description}
\item {\bf Hypothesis 1.} Users speaking on a topic expose their local
  communities to the topic. The topic grows as it spreads to more
  communities, and not just more users in the same community.
\item {\bf Hypothesis 2.} The merging of several local communities
  talking on a viral topic leads to the topic attaining its
  peaks. 
\item {\bf Hypothesis 3.} When a topic goes viral, a node speaking on
  the topic has many of its strong ties speaking on the topic as
  well. Subsequent discussions on this topic stay largely within the
  conglomerate of local communities initially formed.
\end{description}
The crucial part of the story is that it is local communities that
support a topic during its early life. When a number of local
communities gain, and sustain, interest in the topic these communities
begin to merge. To understand this take the following example: if the
community of musicians is excited about a topic and the community of
sports enthusiasts is talking about a topic, if the topic sustains
then sooner or later the musicians who follow sports will talk about
it, causing the communities to merge. The crucial part of this merger
is encapsulated in Hypothesis 3: It {\em does not} take place through
weak long-range links, it happens along strong local links. This
merger of several local communities takes the topic viral. And once it
has gone viral, it is this merged conglomerate of local communities
that continues to discuss the topic (it does not spread significantly
outside them) till it slowly yields to newer topics.

Before we present the evidence in support of our hypotheses we
formalize the notion of local community. We define a {\em lattice
  cluster} of size $m$ centered around node $u$ in a Watts-Strogatz
graph with mean degree $k$ as the set of all nodes that can be reached
from $u$ in $m$ hops or more where the hops are taken along lattice
edges only. The notion of a lattice cluster springs from the
resemblance of $k/2$ levels in a Watts-Strogatz graph with local
communities in a social network. The connectedness of a $k/2$ level
resembles the strong ties within a community sharing homophily. A
common subject of interest holds the nodes in a level together. When a
topic enters a community, it immediately gets passed on within the
community by virtue of the users sharing a common interest. This real
phenomenon manifests itself in the dynamics too. An adoption is
quickly followed by successive copies along the lattice neighbors
owing to their connectedness. The growth along the lattice is hence of
essence to the evolution of the topic, since this spread is rapid and
sustainable.

\paragraph{Hypothesis 1. Gain and sustain local support} 

\begin{figure*}[t]
\centering
\begin{tabular}{cc}
\begin{minipage}{5.8cm}
\epsfig{file = 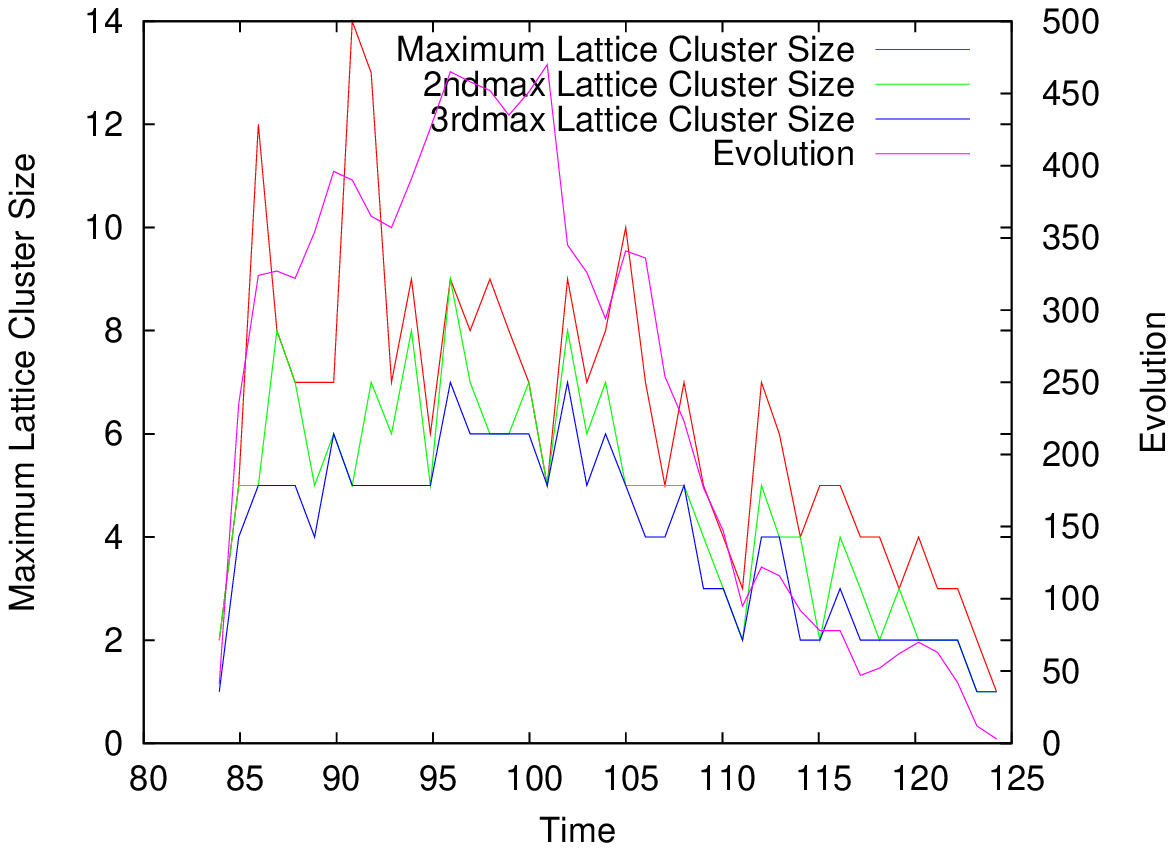, width=5.8cm,height=4cm}
\end{minipage}&
\begin{minipage}{5.8cm}
\epsfig{file = 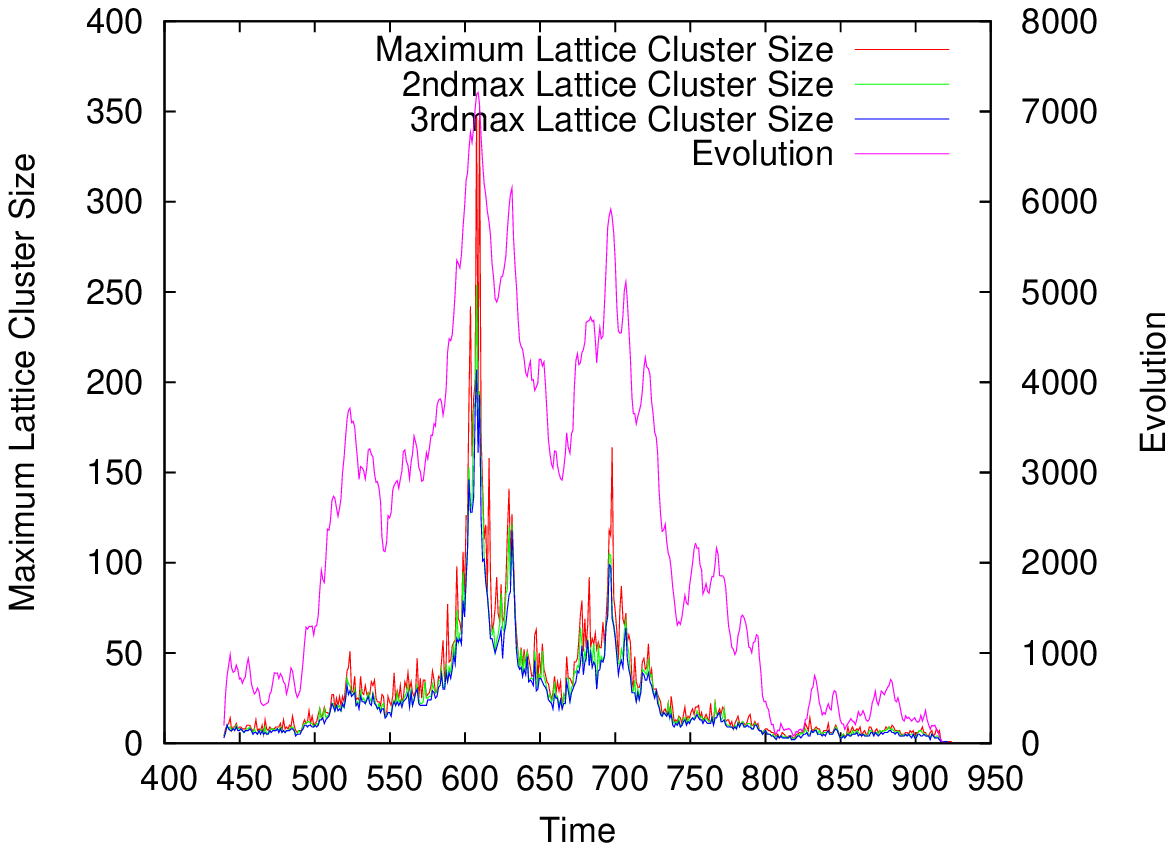, width=5.8cm,height=4cm}
\end{minipage}\\
(a) Non-viral topic &(b) Viral topic\\
\end{tabular}
\caption {Graphs depicting evolution of lattice cluster diameter for a non-viral and a viral topic}
\label{fig:lattice_cluster_evolution}
\end{figure*}

Figure~\ref{fig:lattice_cluster_evolution} shows the evolution of a
topic along with the evolutions of the three largest ranking lattice
clusters. One can see the early life of the topic being characterized
by many lattice clusters of small sizes. Had the topic grown by virtue
of local communities buzzing with the topic and the neighbouring
communities then catching it on from them, one would have seen a large
sized lattice cluster followed by smaller ones early in the evolution
of the topic. But, the comparable sizes of the three top ranking
lattice clusters, is indicative of a spread to adjoining lattice
clusters without saturation of the initiating one.

\begin{figure*}[t]
\centering
\begin{tabular}{cc}
\begin{minipage}{5.8cm}
\epsfig{file = 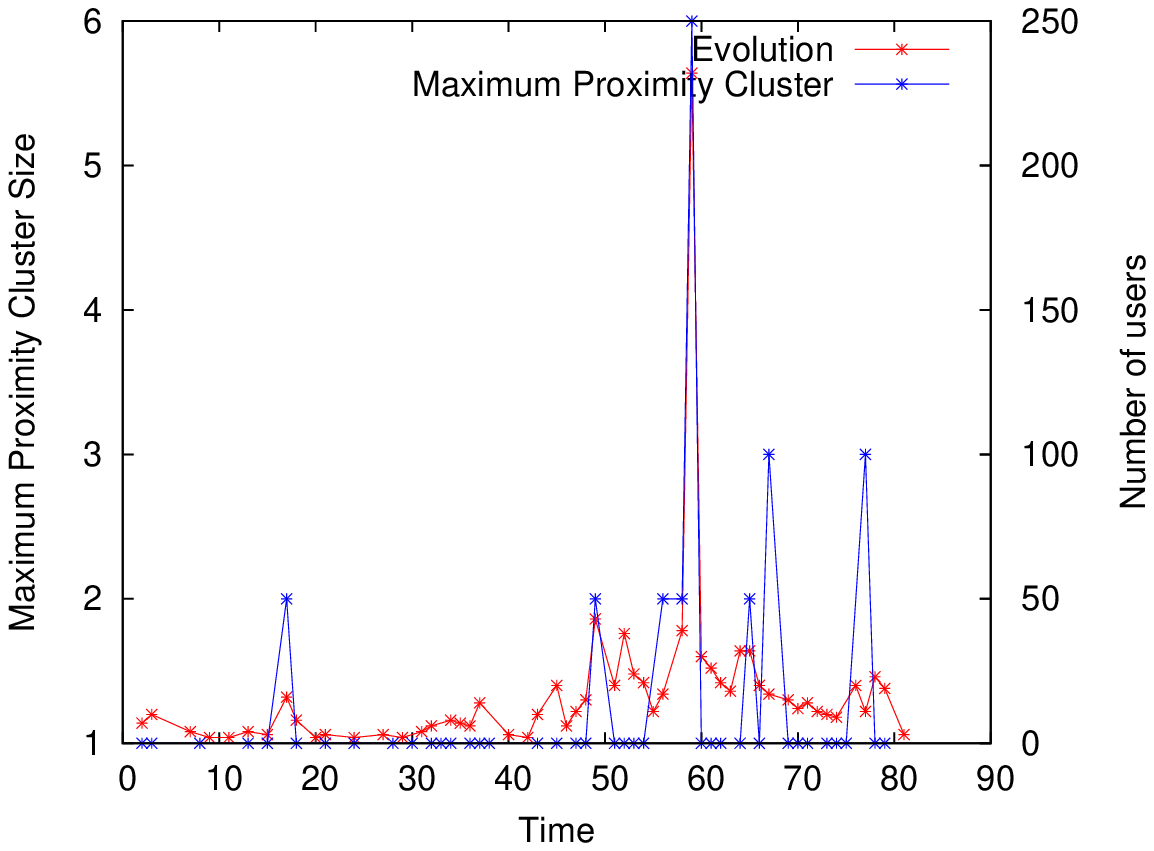, width=5.8cm,height=4cm}
\end{minipage}&
\begin{minipage}{5.8cm}
\epsfig{file = 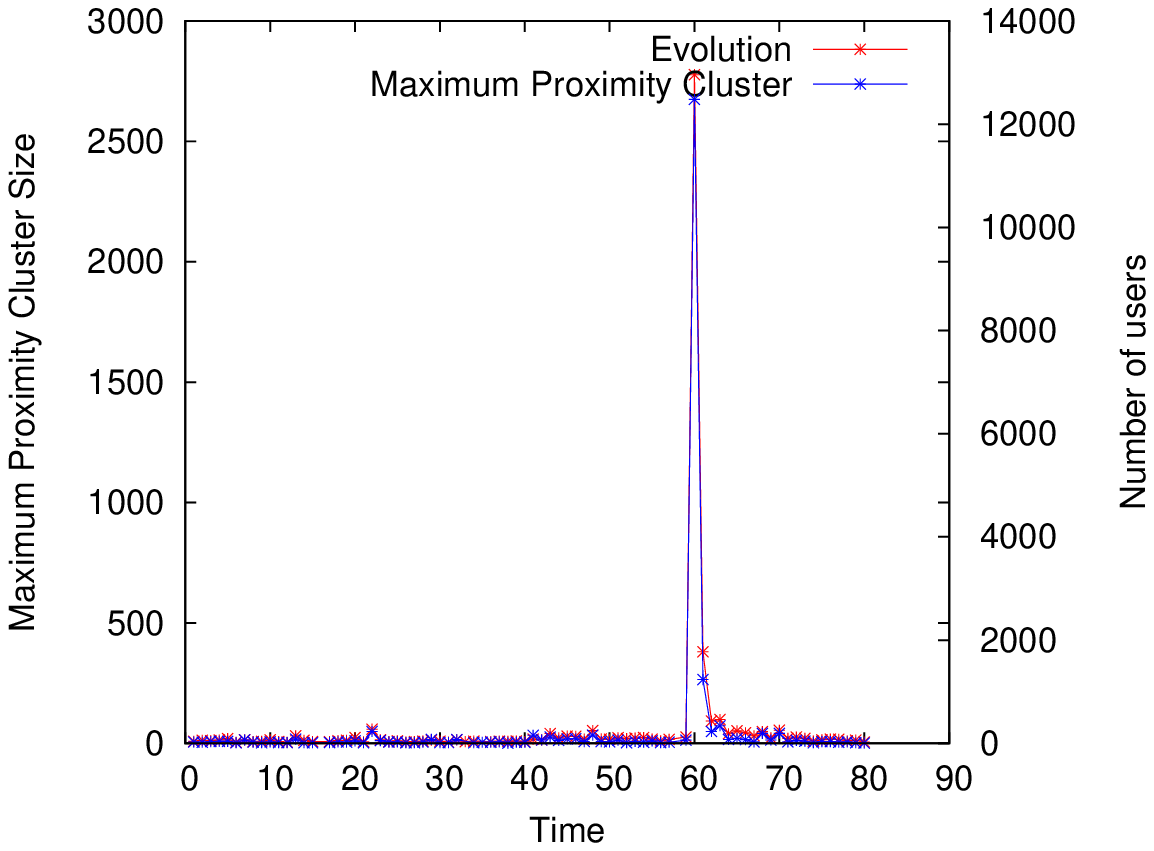, width=5.8cm,height=4cm}
\end{minipage}\\
(a)Non-viral(CARY GRANT) &(b) Viral topic (FRIENDFEED)
\end{tabular}
\caption{Proximity cluster size: Real}
\label{fig:proximity}
\vspace{-.1in}
\end{figure*}

\paragraph{Hypothesis 2. Merging localities lead to virality}
When two such strings along the lattice collide, the nodes in the
levels around the point of collision begin to fill in quickly. We say
two lattice clusters have merged when the $k/2$ levels between them
have at least one node speaking on the topic. The strings then
coalesce to form one big lattice cluster. In the next few time steps,
this burst of weight of the topic begins to spread to $k/2$ levels on
either side of the point of collision on the lattice. As the lattice
cluster begins to fill in, more and more nodes within the cluster
begin to hear about the topic and speak on it. This phenomenon leads
to the topic peaking sharply in its evolution, suppressing other
topics. When the burst of weight has spread as far as it could, the
lattice cluster saturates in size. The massive weight of the topic
within this lattice cluster and the connectedness of the cluster
enable nodes to still continue copying the topic from each other,
speaking intermittently on the topic and enabling it to sustain its
peak for a finite duration.

Figure \ref{fig:lattice_cluster_evolution} shows the difference in the
sizes of the lattice clusters for a viral and a non-viral
topic. Further, for a viral topic, the size of the lattice cluster
shoots up when the topic peaks in its evolution, showing a clear
correlation between the two. The evolution has a peak $\sim 7,000$
while the largest lattice cluster has a peak $\sim 350$, showing that
all nodes speaking on the topic at that time are not in one giant
lattice cluster. There are several lattice clusters of varying sizes
extant at the same time. However, the merging of a few clusters to
form a giant component is in sync with the peak in the evolution of
the topic. 

To get a feel for lattice-clusters in real Twitter data, we defined a
notion of proximity. Two nodes $u$ and $v$, in the topic subgraph,
have a proximity value associated with them, which is measured as:
\begin{center}
$P(u,v)=\dfrac{|\tau(u)|\cap |\tau(v)|}{|\tau(u)|\cup |\tau(v)|}$
\end{center}
where $\tau(u)$ denotes the set of nodes in the neighborhood of
$u$. Thus, more the fraction of common neighbors, more the value of
proximity between a pair of nodes. We put a threshold $T_{p}$ on the
proximity value and consider only the nodes with $P(u,v)>T_{p}$ as
being lattice neighbours of each other.

Using the notion of proximity, we define a proximity cluster ($PC$) at
time $t$ as a connected component in the topic subgraph, talking on
the topic, which contains a node $u$, only if $\exists$ at least a $v
\in PC$ such that $P(u,v)>T_{p}$.  In Figure~\ref{fig:proximity}, we
plot graphs for a viral and a non-viral topic from the real data,
showing how the maximum proximity cluster size varies with the
evolution of the topic. The plot shows a clear correlation between the
evolution of the largest lattice cluster and the evolution of the
topic. A peak in the topic evolution is synchronously accompanied by a
peak in the size of the largest lattice cluster, thus validating our
hypothesis.

To elucidate the phenomenon of merging in real data, we look at the
evolution of conductance of the topic subgraph. We use the following
definition of conductance. The conductance $\phi(S)$ of a subset of
nodes $S$ of graph $G=(V,E)$ is defined as the ratio of the edges
going out from $S$ to all the edges originating from $S$.
\[ \phi(S) = \frac{|\{(u, v) \in E[G] : u\in S, v \in V\setminus
    S\}|}{|\{(u,v) \in E[G] : u\in S\}|}.\]

\begin{figure*}[t]
\centering
\begin{tabular}{cc}
\begin{minipage}{5.8cm}
\epsfig{file = 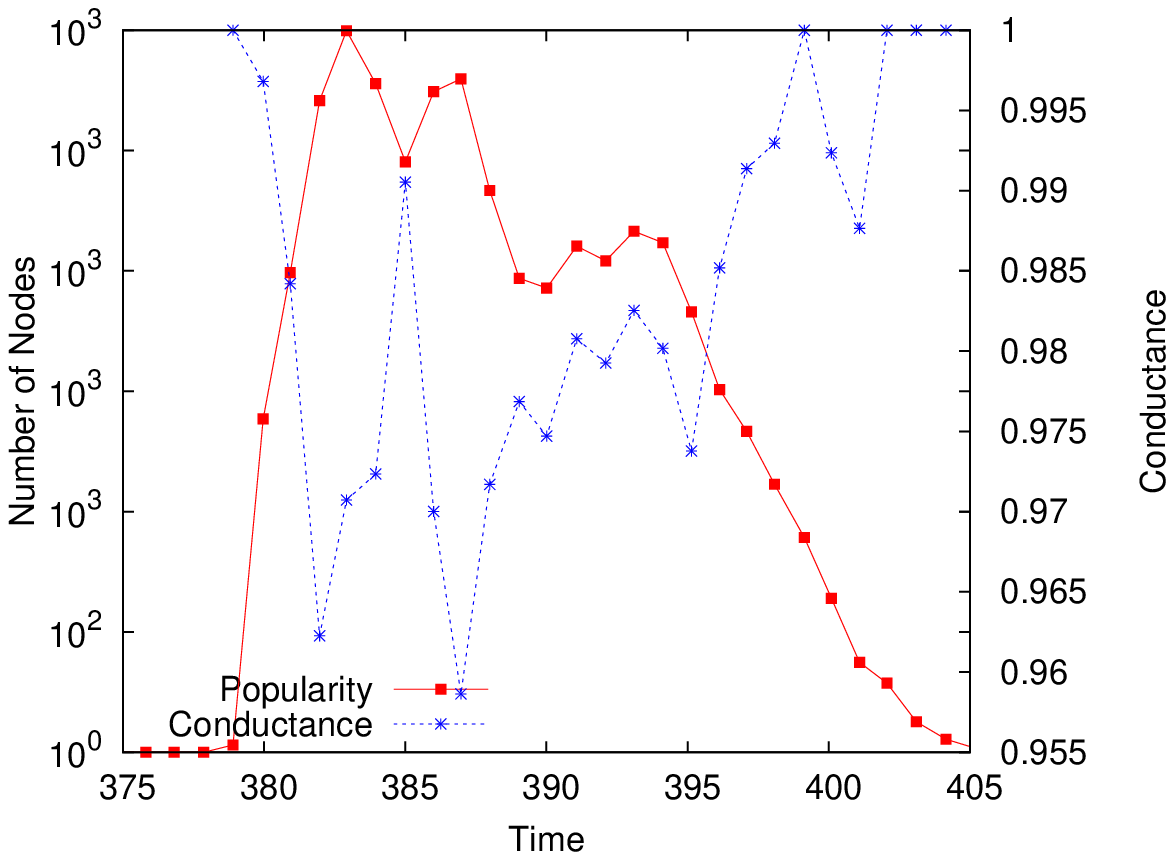, width=5.8cm,height=4cm}
\end{minipage}&
\begin{minipage}{5.8cm}
\epsfig{file = 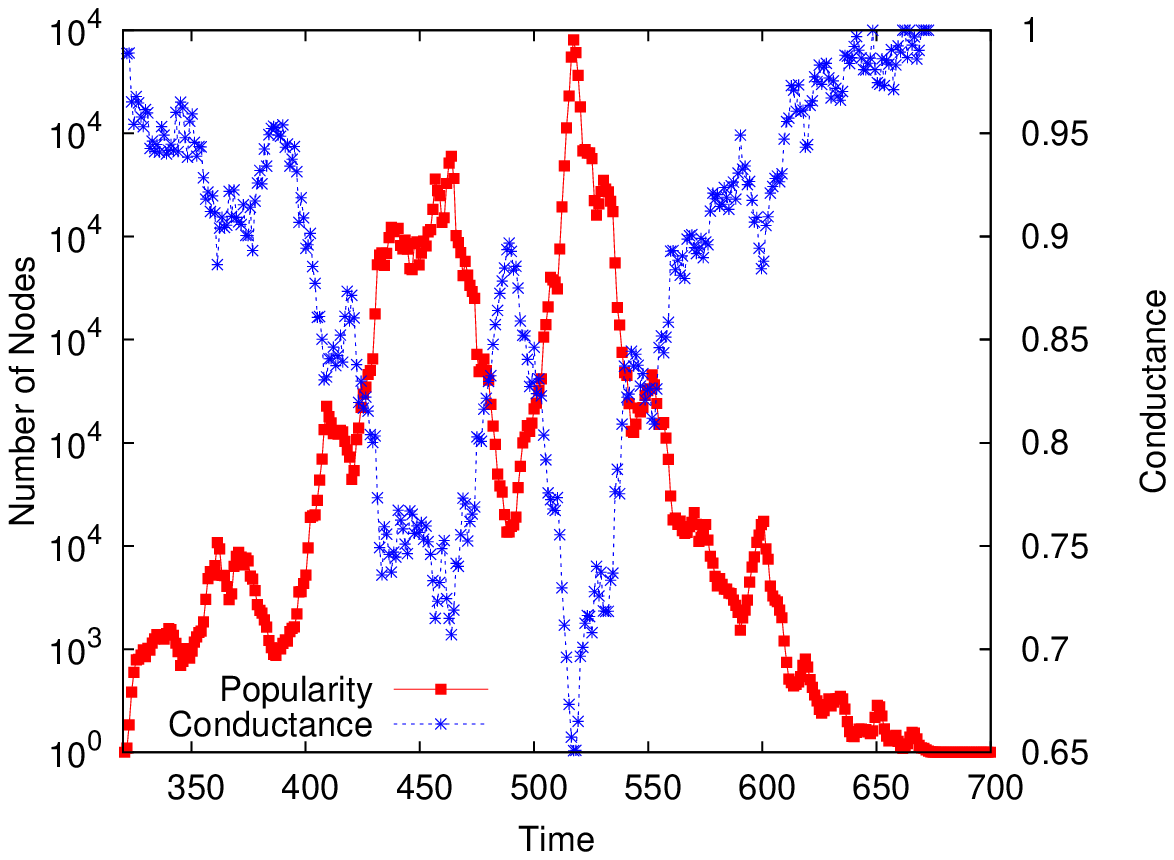, width=5.8cm,height=4cm}
\end{minipage}\\
(a) Simulation: Non-viral &(b) Simulation: Viral\\
\begin{minipage}{5.8cm}
\epsfig{file = 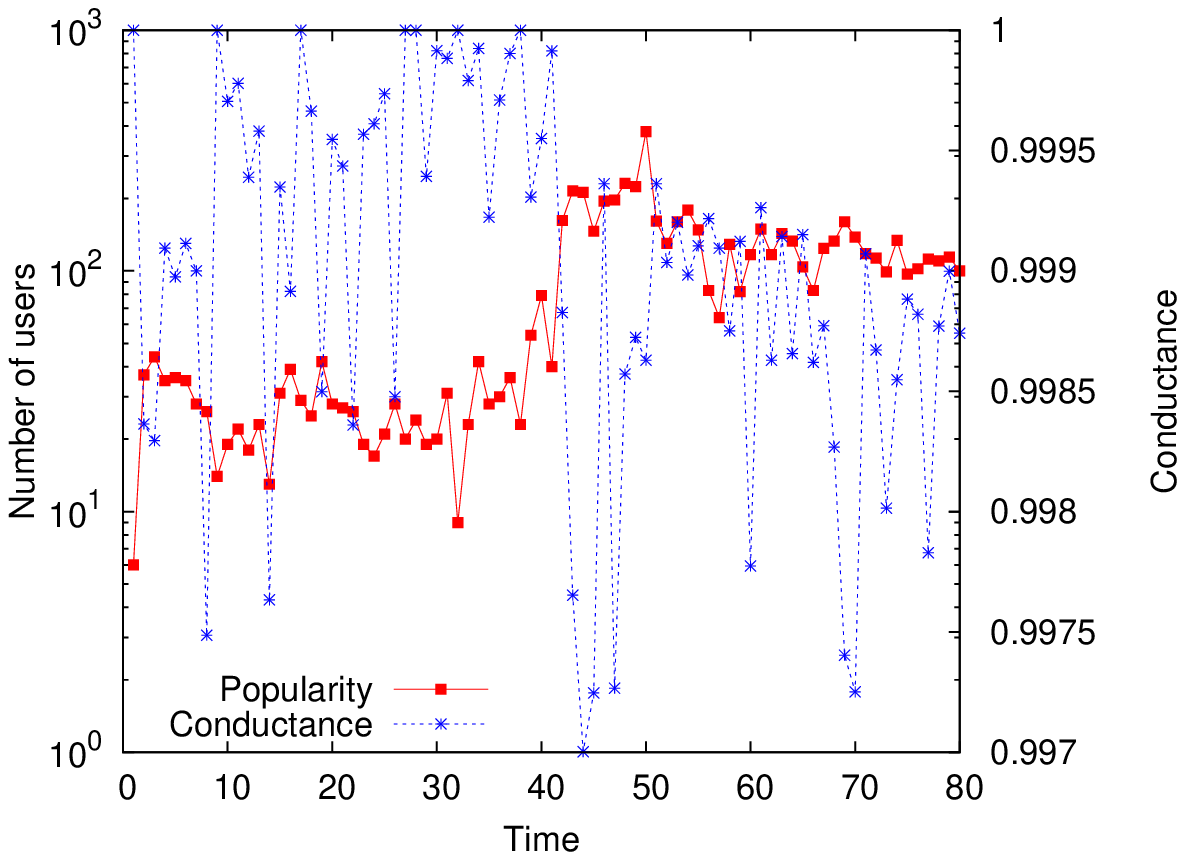, width=5.8cm,height=4cm}
\end{minipage}&
\begin{minipage}{5.8cm}
\epsfig{file = 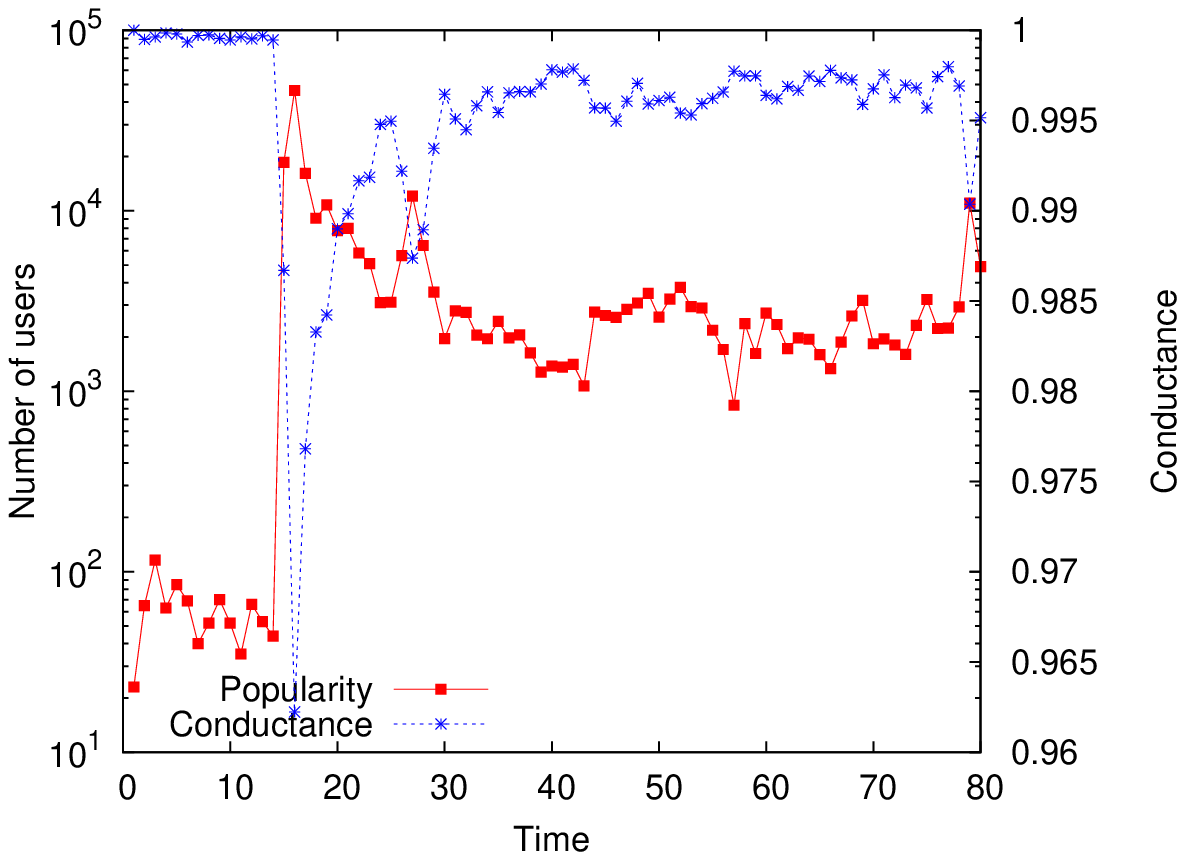, width=5.8cm,height=4cm}
\end{minipage}\\
(c) Real: Non-viral &(d) Real: Viral\\
 (CAMBRIDGE) & (MICHAEL JACKSON)
\end{tabular}
\caption{Conductance: Simulation vs Real}
\label{fig:conductance}
\vspace{-.1in}
\end{figure*}
Figure~\ref{fig:conductance} shows the conductance graphs for
simulation and real data for a viral and a non-viral topic. The
general pattern in the evolution of conductance is similar in the real
and simulated curves. With the merging of lattice clusters and with
more and more nodes within the cluster speaking on the topic, we would
expect the number of edges within the cluster to increase and the
number of outgoing edges to decrease. Consequently we would expect the
conductance value to dip when the clusters
merge. Figure~\ref{fig:conductance} shows the conductance value
dipping when the topic peaks, thereby suggesting merging of
clusters. If we look at the magnitudes of conductance values, we find
that for the viral case, the values drop to as low as
$0.65$(simulations) and $0.96$ (real-data), indicating the formation
of strongly connected components at the peak of topic evolution. On
the other hand for the non-viral cases, we have much higher values.

\paragraph{Hypothesis 3. Virality is sustained in the merged cluster.}
For a particular topic we observed the value of the local weight of
that topic in the neighborhood of each node at each time that the node
speaks on that topic. i.e. if a node $u$ at time $t$ spoke on topic
$i$ and the weight of topic $i$ in the neighborhood of $u$ was
$w_t^u(i) = w$, say, then we noted this as one occurrence of the value
$w$ and plotted a histogram of the frequency of different weight
values.
\begin{figure*}
\centering
\begin{tabular}{cc}
\begin{minipage}{5.8cm}
\epsfig{file = 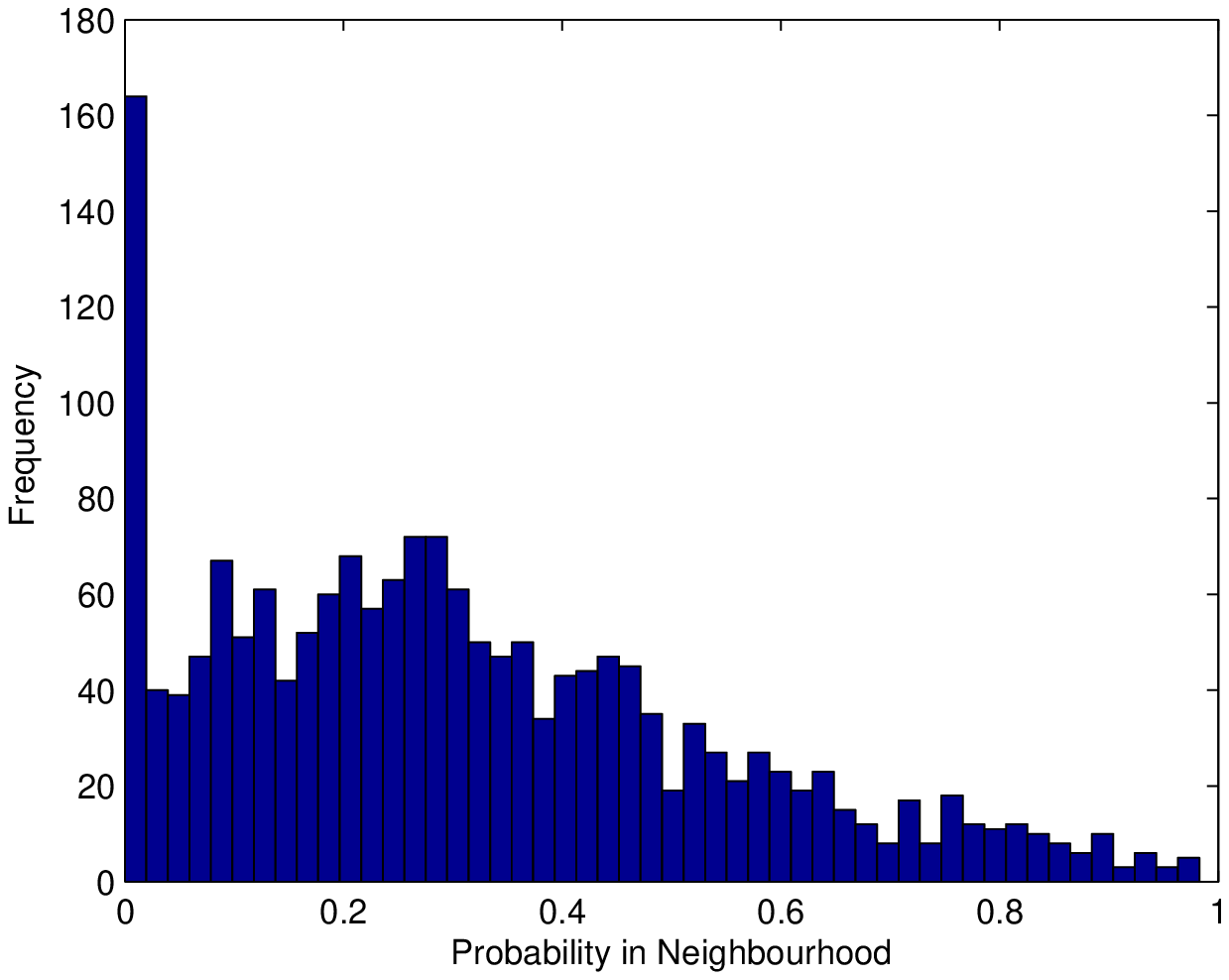, width=5.8cm,height=4cm}
\end{minipage} &
\begin{minipage}{5.8cm}
\epsfig{file = 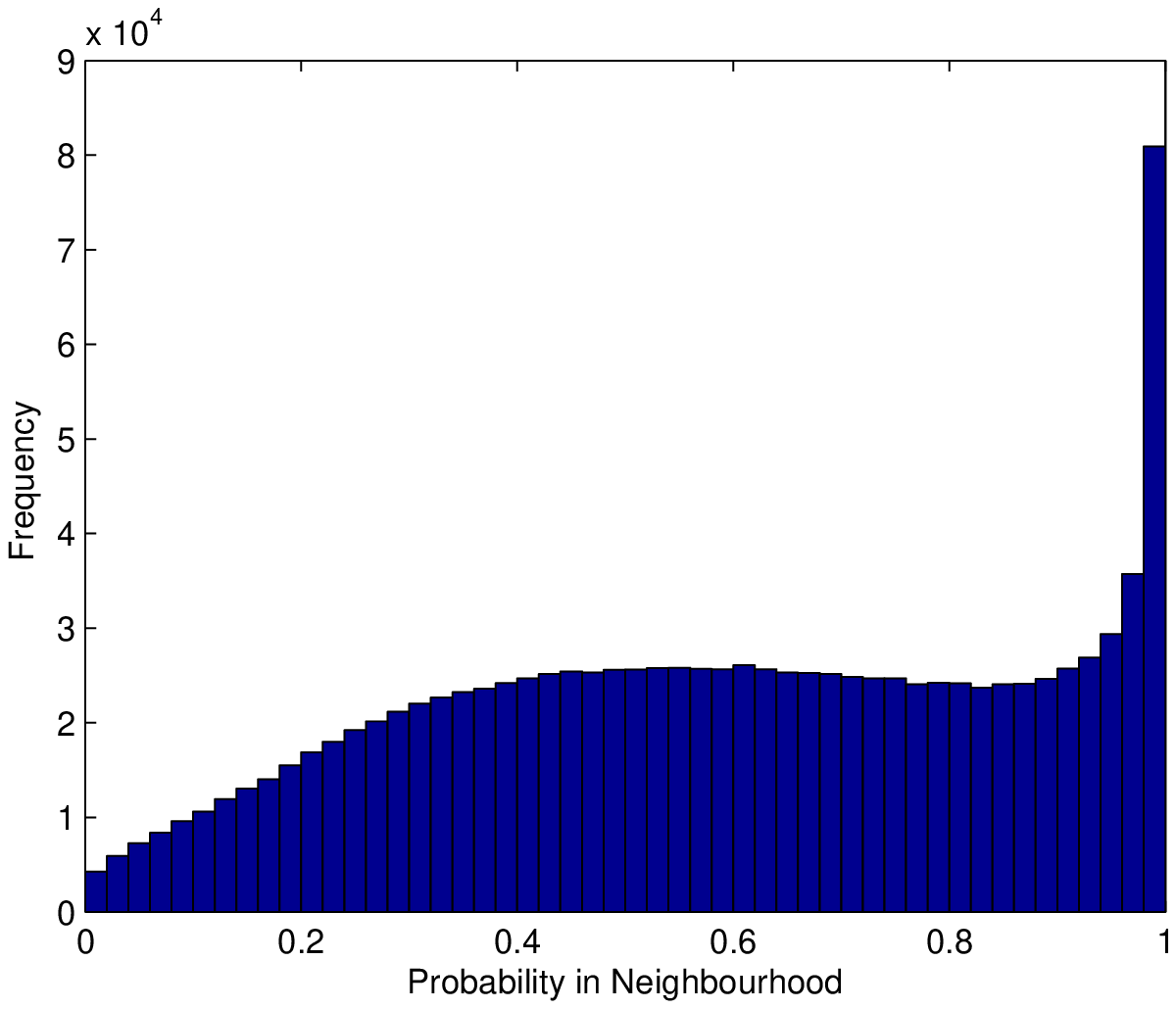, width=5.8cm,height=4cm}
\end{minipage}\\
(a) Non-viral topic & (b) Viral topic\\
\end{tabular}
\caption {Local weight histogram (10,000 nodes)}
\label{fig:prob_histogram}
\end{figure*}
In Figure~\ref{fig:prob_histogram} we show the histograms thus
obtained for a viral topic and a non-viral topic from the viral regime
on a network of size 10000.  Firstly, we observe the difference in the
peak heights of the histograms for a viral and a non-viral topic. This
is a result of a viral topic being spoken about many number of times
as compared to a non-viral topic. For a non-viral topic, the local
weights that are achieved maximum number of times hover around 0.2 to
0.4. Beyond 0.5, the peaks begin to dip. Since achieving a high local
weight is equivalent to more number of neighbors of the node talking
on the topic in quick succession, the dip shows that such scenarios
for a non-viral topic are rare. For a viral topic, however, we observe
a different pattern with the histograms peaking as we move closer to
$1$. This shows that for a viral topic, the scenarios where a node has
most of its neighbors talking on the topic at quick successions are
huge in number. In fact, it is this scenario of multiple neighbors
talking simultaneously on the topic which makes the neighboring nodes
adopt the topic with greater probability driving it viral. The peak in
the local weight for the topic leads to it being chosen with greater
probability which acts as a feedback making the topic go further viral
and also sustain the virality.
\begin{figure}
\centering
\epsfig{file = 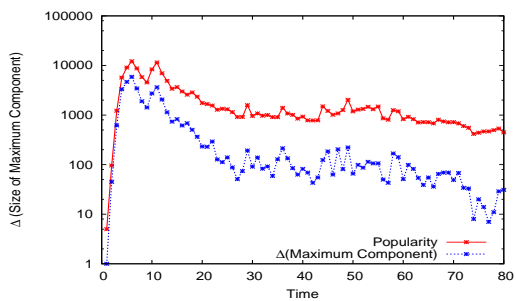, width=7cm,height=4cm}
\caption {Change in number of users talking about IRANELECTION}
\label{fig:iranelection_delta}
\end{figure}
The real data shows a related phenomenon. In
Figure~\ref{fig:iranelection_delta} we see that once the topic
``IRANELECTION'' has reached its peak (of the order of $10^5$), the
change in the number of users is bounded in the order of a few hundred
a day, often falling to much lower values. Hence virality, if
sustained, is generally sustained by largely the same user base that
caused it to occur in the first place, thereby making us feel that it is
 at the point of its emergence that a viral topic garners most of
the support it is ever going to get.

\paragraph{Discussion}
The phenomenon of lattice cluster merging exhibits how a topic which
is popular in a small group of users sharing homophily need not take
over the entire network. If we consider the hops of $k/2$ as being a
gradual shift in the subject binding these users closely to each
other, the spread of the topic even when it is young shows the need
for a few people in different related genres to speak on the topic
simultaneously. These people influence more users to talk and the
lattice cluster spreads. When the lattice clusters from two different
genres collide, the intermediary users are encouraged to copy the
topic owing to homophily on either side. Both these genres attack the
intermediary nodes which then begin to spread the buzz. The copying
then sees both the distinct genres filling up as more and more
hardcore users of the genre copy the topic. The quick filling in shows
why most of the users do not seem to hear about the topic at all and
then hear a lot about it suddenly and from everywhere, everywhere
being a sign of the different genres that the node is a part of.

\section{Implications and future directions}
\label{sec:conclusions}

Much has been said in the last two decades about ``six degrees of
separation'' and ``small worlds'' and it is true that the world is
more closely linked today than it has ever been. But our results show
that virality is not a function of the social network's ability to
bring faraway things closer. It is, in fact, achieved when many
smaller homophilic groups discover an interest in a certain topic. The
overlapping of this interest over a large number of such groups leads
to virality. This is, in some sense, an organic narrative of growing
popularity, a narrative that those marketing executives who spend time
poring over the early sales data of The Girl With the Dragon Tattoo
perhaps understand intrinsically. For those of a less commercial bent,
those who want to change the world, also there is a lesson in our
findings. That lesson is if enough people act locally in a large
enough number of localities, there is a global impact. And a sobering
thought is that no matter how intensely one community feels about some
topic, unless other communities also develop that interest, the topic
will die in small disconnected clusters.

The main focus of this work has been on modeling the processes of
diffusion although we have adduced results from a small measurement
study to back up our claim that our model is a real model and that the
artifacts we find in it have echoes in the real world. One prominent
research direction to follow is a large-scale model-driven measurement
study. Another direction that remains is to generalize the model by
allowing users to act at different times, perhaps act in correlated
times rather than independent time, to repose different levels of
trust in different neighbors, to view different classes of topics with
different levels of interest. In the current work we sacrificed this
rich diversity of the topic diffusion setting in order to define a
model that was easier to handle while being sufficiently general as to
describe a coarser level of behavior, but if a more nuanced view is
taken then a richer model emerges. In such a case further measurement
could also help to instantiate different parameter classes.

\end{document}